\documentclass[a4paper,fleqn,usenatbib]{mnras}

\usepackage{newtxtext,newtxmath}

\usepackage[T1]{fontenc}
\usepackage{ae,aecompl}
\usepackage{siunitx}

\usepackage{graphicx}	
\usepackage{amssymb}	
\usepackage{pifont}
\usepackage{makecell}
\usepackage{booktabs}
\usepackage{multirow}
\usepackage{mathtools}
\defcitealias{CC1985}{CC85}
\defcitealias{Silich2004}{S04}
\defcitealias{Thompson2015}{T15}
\defcitealias{Sharma2013}{S13}
\defcitealias{Ipavich1975}{Ip75}

\title[Hydrodynamical study of galactic outflows]{A hydrodynamical study of outflows in starburst galaxies with different driving mechanisms}

\author[B. P. Brian Yu et al.]{
B. P. Brian Yu$^{1,2}$\thanks{E-mail: brian.yu.16@ucl.ac.uk (BPBY), ellis.owen.12@ucl.ac.uk (ERO), kinwah.wu@ucl.ac.uk (KW), iferreras@iac.es (IF)},
Ellis R. Owen$^{1,2}$, Kinwah Wu$^{1,3}$ and Ignacio Ferreras$^{4,5,6}$ \\
$^{1}$ Mullard Space Science Laboratory, University College London, Holmbury St Mary, Dorking, Surrey, RH5 6NT, UK \\ 
$^{2}$ Institute of Astronomy, Department of Physics, National Tsing Hua University, Hsinchu, Taiwan (ROC) \\
$^3$ Perimeter Institute, 31 Caroline St.~N., Waterloo, Ontario, N2L 2Y5, Canada\\
$^{4}$ Department of Physics and Astronomy, University College London,
Gower Street, London WC1E 6BT, UK\\
$^{5}$ Instituto de Astrof\'isica de Canarias, C/V\'ia L\'actea, s/n, E38205 La Laguna, Tenerife, Spain\\
$^6$ Departamento de Astrof{\'i}sica, Universidad de La Laguna (ULL), E-38206 La Laguna, Tenerife, Spain
}

\date{Accepted XXX. Received YYY; in original form ZZZ}

\pubyear{2019}

\begin{document}
\label{firstpage}
\pagerange{\pageref{firstpage}--\pageref{lastpage}}
\maketitle

\begin{abstract}
Outflows from starburst galaxies can be driven by thermal pressure, radiation and cosmic rays. We present an analytic phenomenological model that accounts for these contributions simultaneously to investigate their effects on the hydrodynamical properties of outflows. We assess the impact of energy injection, wind opacity, magnetic field strength and the mass of the host galaxy on flow velocity, temperature, density and pressure profiles. For an M82-like wind, a thermally-dominated driving mechanism is found to deliver the fastest and hottest wind. Radiation-driven winds in typical starburst-galaxy configurations are unable to attain the higher flow velocities and temperatures associated with thermal and cosmic ray-driven systems, leading to higher wind densities which would be more susceptible to cooling and fragmentation at lower altitudes. High opacity winds are more sensitive to radiative driving, but terminal flow velocities are still lower than those achieved by other driving mechanisms at realistic opacities. We demonstrate that variations in the outflow magnetic field can influence its coupling with cosmic rays, where stronger fields enable greater streaming but less driving near the base of the flow, instead with cosmic rays redirecting their driving impact to higher altitudes. The gravitational potential is less important in M82-like wind configurations, and substantial variations in the flow profiles only emerge at high altitude in massive haloes. This model offers a more generalised approach to examine the large scale hydrodynamical properties for a wide variety of starburst galaxies.
\end{abstract}

\begin{keywords}
ISM: jets and outflows -- galaxies: starburst -- hydrodynamics -- cosmic rays -- radiation: dynamics
\end{keywords}



\section{Introduction}

Galactic outflows are present in star-forming galaxies 
  and have been observed in neraby starburst galaxies, e.g.  Arp 220, M82 and NGC 253, 
  and young galaxies further afield 
  \citep{Frye2002ApJ, Ajiki2002ApJ, Benitez2002book, Rupke2005ApJS-a, 
        Rupke2005ApJS-b, Bordoloi2011ApJ, Arribas2014A&A}. 
These outflows generally have a bi-conical structure, 
  directed along the minor axis of their host galaxy, which governs the path of least resistance encountered by an otherwise spherical wind~\citep{Veilleux2005},   and
  they are powered by starburst activity of the host galaxy. Outflow velocities can vary greatly, 
   depending on the underlying driving mechanism of the wind. They have been measured from a few hundred km s$^{-1}$ in most cases, 
   rising to a few thousand km s$^{-1}$ in certain extreme examples \citep{Cecil2002RMxAACS, Rupke2005ApJS-b, Rubin2014ApJ}, while cosmic ray-driven cold winds could be as slow as just a few tens of km s$^{-1}$~\citep{Samui2010, Uhlig2012MNRAS, Samui2018}.

Galactic outflows are a multi-phase, multi-component  media~\citep[see][]{Ohyama2002PASJ,Strickland2002,Melioli2013,Martin-Fernandez2016}, comprised of cool semi-ionised clumps of gas 
(of temperature $T_{\rm c} \sim 10^2-10^4\,{\rm K}$ -- 
 see~\citealt{Strickland1997A&A, Lehnert1999ApJ})
 entrained within a hot ($T_{\rm h} \sim 10^7$~K
 ~\citealt{McKeith1995A&A, Shopbell1998}), low-density X-ray emitting fluid which may extend to altitudes of several kpc~\citep{Strickland2000AJ, Cecil2002ApJ, Cecil2002RMxAACS}. 
 Above this, there is a cap
(in M82, this is observed at 11.6 kpc -- see~\citealt{Devine1999, Tsuru2007}), 
with the full outflow structure extending to tens of kpc~\citep[see][]{Veilleux2005, Bland-Hawthorn2007APSS, Bordoloi2011ApJ, Martin2013ApJ, Rubin2014ApJ, Bordoloi2016MNRAS}.
 In the hot fluid region of an outflow, cooling processes are presumably important and would have a significant impact on the dynamics and evolution of the flow~\citep{Heckman2003}. In winds of intensely star-forming galaxies, this can lead to substantial variation in the thermal properties throughout the flow (see e.g. the profiles shown in~\citealt{CC1985}). Processes including ionisation and mechanical shock heating are also present alongside these cooling mechanisms~\citep{Hoopes2003}.
 
 
The physical origins and the driving mechanisms of galactic outflows 
  have remained unsettled 
  since the first discovery of an outflow in M82 \citep{Lynds1963}. 
While they are fuelled by star-formation (predominantly by the resulting supernova winds), outflow and galaxy properties have complicated inter-dependencies. The star formation rate (SFR) determines  the intensity of the outflow, but the outflow itself may hamper star-formation \citep{Veilleux2005}; galactic morphology shapes the path of least resistance followed by an outflow, but the wind can blow away the interstellar medium (ISM) to change the galactic morphology \citep{Cooper2008ApJ}; the galactic mass and metallicity decide how much of the outflow will be bound by the gravitational potential and how quickly the outflow will cool, but the outflow can launch matter (including metals) out of the galaxy \citep{Dave2009}, and the efficiency of mass entrainment in the winds can vary according to various factors \citep{Rupke2005ApJS-b}. Moreover, galaxy properties evolve over redshift~\citep{Mannucci2010}: galaxies in the early Universe are bluer \citep{Madau1996} and smaller \citep{Dickinson2003} than those today, and this would have an effect on outflow properties and characteristics~\citep[e.g., see][]{Sugahara2019}.
The driving mechanism(s) of an outflow are governed by the properties of their host galaxy: while radiation pressure may be important in a metal rich galaxy, thermal pressure may dominate in a metal-poor galaxy (which would also cool significantly more slowly).
As such, a broad range of hydrodynamical (HD) models have been developed
to account for different driving mechanisms. Early models invoked thermally-driven flows, where the confluence and adiabatic expansion of hot winds forms an outflow~\citep{CC1985, Silich2004}. Later models explored the role of radiation pressure, particularly on dusty winds -- see \citep{Dijkstra2008MNRAS, Nath2009MNRAS, Sharma2013, Thompson2015} as well as cosmic rays~\citep{Ipavich1975, Samui2010}. 
In this paper, we present a generalised model that takes all three  of these driving mechanisms into account. In section \ref{sec:model}, the stationary solutions of each of the individual HD models are explored, 
together with our generalised model. 
The corresponding results (particularly the profiles of the HD quantities)
and their astrophysical implications are discussed in section \ref{sec:compare}. We draw conclusions in section~\ref{sec:conclusion}.

\section{Hydrodynamical Models}
\label{sec:model}

In this paper, we assume that the development of a galactic outflow can be ascribed to three major physical contributors: the thermal content of the gas; radiation pressure; or cosmic rays (hereafter CR). We calculate the stationary solutions of the HD equations under these three potential driving scenarios, together with a generalised prescription that combines all three.
In the following, we neglect the effect of turbulence (i.e. we assume an inviscid flow) for analytical tractability, and a spherically symmetric geometry is adopted throughout.
We outline the original HD models for each of the driving mechanisms in the following, and indicate the modifications we have made, leading to the generalised model which we discuss in section~\ref{sec:mix}.
Unless otherwise stated, we adopt reference model parameters to emulate a system similar to M82, as summarised in table \ref{tab:param}. The regime of validity of our model requires that the ram pressure of the supersonic wind should be higher than the ambient gas pressure. The radial profiles shown throughout this paper extend to 10\,kpc, much less than the virial radius, so that these constraints are fulfilled.



\begin{table}
\centering
\begin{tabular}{*{3}{c}}
\midrule
Parameter & Value & Reference \\
\midrule
\(r_{\rm sb}\) & \(200\ \rm pc\) & \cite{Shopbell1998} \\
\(\dot{M}\) & \(2.6\ M_\odot/\rm yr\) & \cite{Veilleux2005} \\
\(\dot{E}\) & \(4.2\times10^{41}\ \rm erg/s\) & \cite{Veilleux2005}  \\
\(\kappa\) \textsuperscript{\textit{a}} & \(10^4\ \rm cm^2/g\) & \cite{Sharma2013} \\
\(B_0\) \textsuperscript{\textit{b}} & \(50\ \mu \rm G\) & \cite{Klein1988} \\
\(M_{\rm tot}\) \textsuperscript{\textit{c}} & \(5.54\times 10^{11}\ M_\odot\) & \cite{Oehm2017} \\
\(R_{\rm s}\) \textsuperscript{\textit{c}} & \(14.7\ \rm kpc\) & \cite{Oehm2017} \\
\(R_{\rm vir}\) \textsuperscript{\textit{c}} & \(164\ \rm kpc\) & \cite{Oehm2017} \\
\midrule 
\end{tabular}
\caption{A list of reference parameters for a starburst system representative of M82, as used for the baseline model in our HD models. Notes: \newline
\textsuperscript{\textit{a}} \(\kappa\) is the mean opacity of the wind, averaged over all frequencies. \newline
\textsuperscript{\textit{b}} \(B_0\) is the maximum galactic magnetic field strength (external to the wind). \newline
\textsuperscript{\textit{c}} \(M_{\rm tot}\), \(R_{\rm s}\) and \(R_{\rm vir}\) are the parameters for the \citealt{Navarro1996} (NFW) dark matter profile.}
\label{tab:param}
\end{table}

\subsection{Thermal outflows}\label{sec:thermal}


\subsubsection{Initial model}
\label{sec:CC85}

\citet[][hereafter \citetalias{CC1985}]{CC1985} developed an analytic model for thermally-driven outflows, and applied it to describe the galactic wind from the starburst galaxy M82. 
In this prescription, the HD equations are written as
\begin{gather}
\label{eq:CCmass}
\frac{1}{r^2}\frac{\rm d}{{\rm d}r}\left(\rho vr^2\right)=q \ , \\
\label{eq:CCmomentum}
\rho v\frac{{\rm d}v}{{\rm d}r} = -\frac{{\rm d}P}{{\rm d}r}-qv \ , \\
\label{eq:CCenergy}
\frac{1}{r^2}\frac{\rm d}{{\rm d}r}\left\{\rho vr^2\left(\frac{v^2}{2}+\frac{\gamma_{\rm g}}{\gamma_{\rm g}-1}\frac{P}{\rho}\right)\right\}=Q_{\rm th} \ , 
\end{gather}
where \(r\), \(v\), \(P\) and \(\rho\) are radius, velocity, gas pressure and density respectively. The flows of mass, momentum and energy are governed by equations \ref{eq:CCmass}, \ref{eq:CCmomentum} and \ref{eq:CCenergy}, respectively, and gravity is assumed to be negligible\footnote{This was justified by \citetalias{CC1985} in that the terminal outflow velocity exceeds the escape velocity by an order of magnitude. This is also reinforced by the results from section \ref{sec:G}.}.
The adiabatic index \(\gamma_{\rm g}=5/3\) indicates that the outflow expands freely as the thermal energy is converted to bulk kinetic energy. The injection rates of mass and energy are assumed to be spatially uniform up to a starburst radius \(r_{\rm sb}\), outside of which they are set to zero, i.e.
\begin{gather}
\label{eq:qm}
q=\begin{dcases}
\frac{3\dot{M}}{4\pi {r_{\rm sb}}^3}&\text{if }r< r_{\rm sb} \\
0&\text{if }r\geq r_{\rm sb}
\end{dcases} \ , \\
\label{eq:qe}
Q_{\rm th}= \begin{dcases}
\frac{3\dot{E}}{4\pi {r_{\rm sb}}^3}&\text{if }r< r_{\rm sb} \\
0&\text{if }r\geq r_{\rm sb}
\end{dcases} \ , 
\end{gather}
where \(\dot{M}\) and \(\dot{E}\) are the mass and energy injection rates. As \(r\to\infty\), all thermal energy is converted to bulk kinetic energy, and the terminal velocity of the flow is equal to \({v_\infty}^2=2\dot{E}/\dot{M}=2Q_{\rm th}/q\). By substituting equation \ref{eq:CCmass} into \ref{eq:CCenergy}, the sound speed \(c_{\rm s}\) follows as
\begin{equation}
\label{eq:CCcs}
{c_{\rm s}}^2=\frac{\gamma_{\rm g}P}{\rho}=\frac{\gamma_{\rm g}-1}{2}\left({v_\infty}^2-v^2\right) \ . 
\end{equation}

The position ($r$) and velocity ($v$) can be written in dimensionless form, as \(x=r/r_{\rm sb}\) and \(u=v/v_\infty\) and the derivative \({\rm d}u/{\rm d}x\) follows \citep[by substituting equation \ref{eq:CCcs} into equation \ref{eq:CCmomentum}, see][]{Canto2000, Rodriguez2007} as
\begin{equation}
\label{eq:CCdvdr}
\frac{{\rm d}u}{{\rm d}x}=\begin{dcases}
\frac{u\left(1+\frac{5\gamma_{\rm g}+1}{\gamma_{\rm g}-1}u^2\right)}{x\left(1-\frac{\gamma_{\rm g}+1}{\gamma_{\rm g}-1}u^2\right)}&\text{if }x\leq 1 \\
\frac{2u\big(1-u^2\big)}{x\left(\frac{\gamma_{\rm g}+1}{\gamma_{\rm g}-1}u^2-1\right)}&\text{if }x\geq 1
\end{dcases} \ , \end{equation}
which may be solved to give \(x(u)\) (note that its inverse function \(u(x)\) cannot be found analytically):
\begin{equation}
\label{eq:CCsolution}
x=\begin{cases}
Au\left(1+\frac{5\gamma_{\rm g}+1}{\gamma_{\rm g}-1}u^2\right)^{-\frac{3\gamma_{\rm g}+1}{5\gamma_{\rm g}+1}}&\text{if } x\leq 1 \\
Bu^{-\frac{1}{2}}\left(1-u^2\right)^{-\frac{1}{2\left(\gamma_{\rm g}-1\right)}}&\text{if } x\geq 1 
\end{cases} \ . \end{equation}

\begin{figure}
\includegraphics[width=\columnwidth]{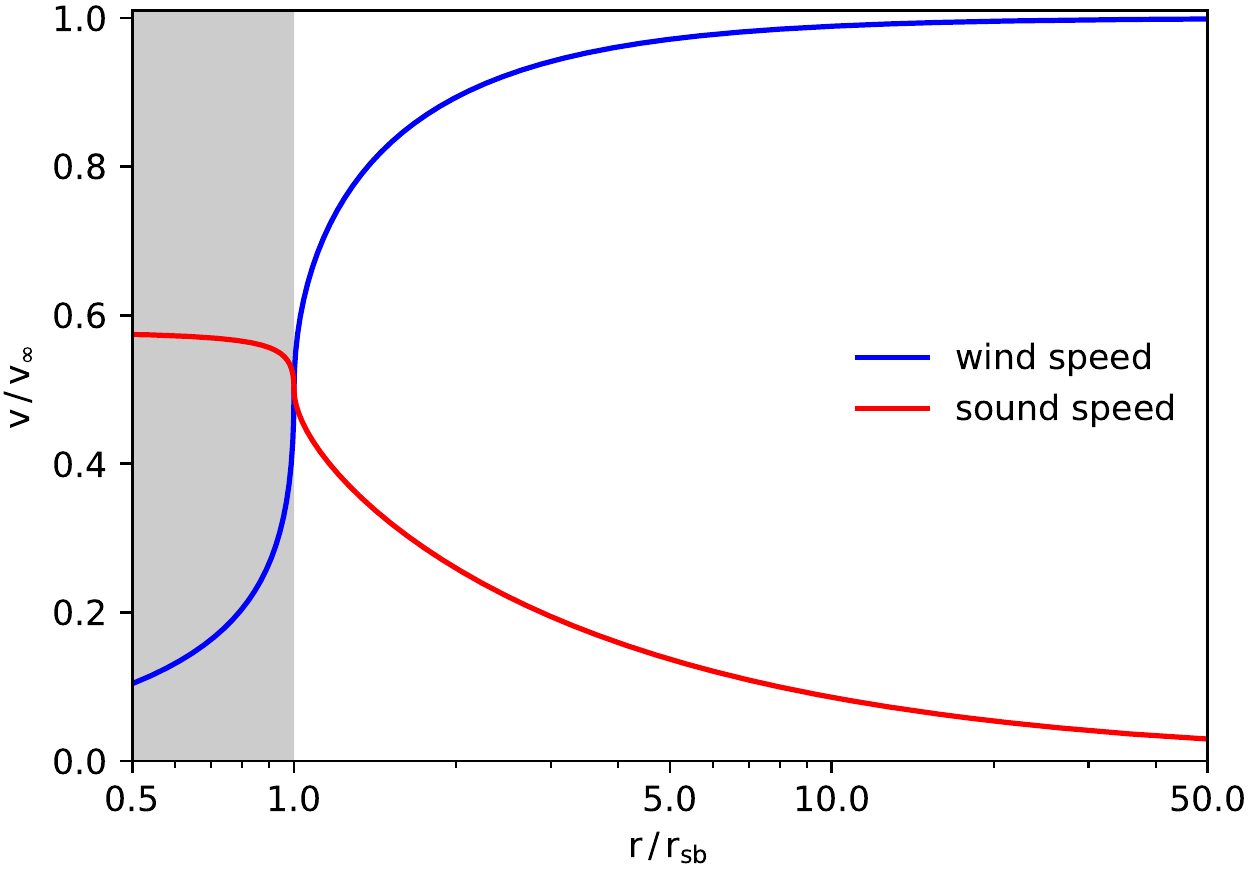}
\caption{Normalised profiles for the wind velocity (blue) and local sound speed (red) in the thermally-driven outflow model. The flow becomes supersonic at the starburst radius, $r_{\rm sb}$. The region inside the starburst nucleus is marked by the shaded area (also in later figures), which is a small region compared to the outflow wind zone. The internal region is of less interest in this work because its structure is not well described by the current HD approach. This is due to complicated details regarding the exact stellar distribution (governing mass/energy injection), turbulence, local ISM flows, CR diffusion \protect\cite{Fujita2018}
and magnetic fields (among other factors).}
\label{fig:CCvelocity}
\end{figure}

The flow velocity must be single-valued for all \(x\). By inspecting the denominator in the two cases of equation \ref{eq:CCdvdr}, the only way to satisfy this is to establish the following boundary condition:
\begin{equation}\label{eq:CCboundary}
v^2\left(r_{\rm sb}\right)
=\left( \frac{\gamma_{\rm g}-1}{\gamma_{\rm g}+1}\right){v_\infty}^2 \ ,
\end{equation}
which can be used to specify the values of the integration constants \(A\) and \(B\) in equation \ref{eq:CCsolution}. 
A consequence of equation \ref{eq:CCboundary} is that \(v=c_{\rm s}\) at \(r_{\rm sb}\) (cf. equation \ref{eq:CCcs}), which
defines a sonic radius (the point at which the outflow becomes supersonic) as $r_{\rm s}=r_{\rm sb}$.
We show the full solution 
of the wind speed 
in Fig.~\ref{fig:CCvelocity} together with the evolution of the sound speed,
which illustrates the outward acceleration of the wind during its adiabatic expansion. 
The step-like nature of the mass injection term \(q\) (equation~\ref{eq:qm}) 
leads to a sharp pressure gradient across the boundary at $r_{\rm sb}$. Beyond this, the wind is no longer required to push against newly injected material 
and is able to accelerate rapidly.
 We note that the values of \(r_{\rm sb}\) and \(v_\infty\) have no influence over the dimensionless velocity profile in Fig~\ref{fig:CCvelocity}, which means that the velocity profile scales directly with the starburst radius \(r_{\rm sb}\) and the injection parameters $q$ and $Q_{\rm th}$.


\subsubsection{Radiative cooling}
\label{sec:radcool}
\begin{figure*}
\includegraphics[width=2\columnwidth]{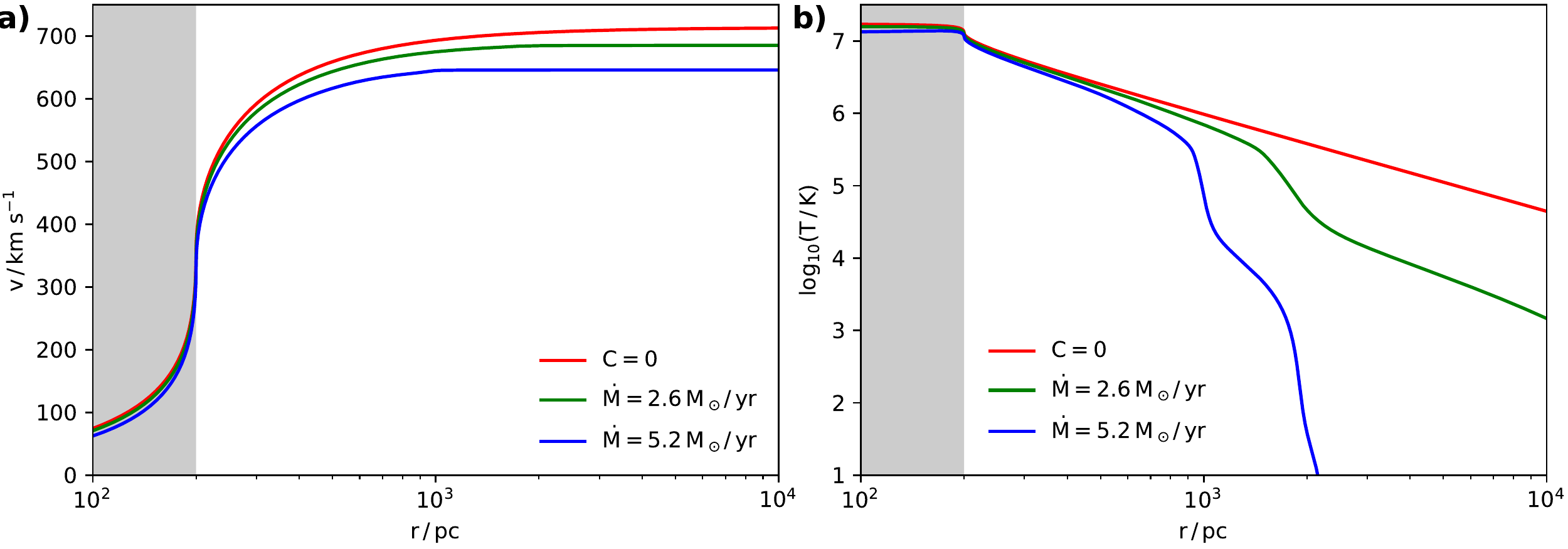}
\caption{Velocity (panel 2a) and temperature (panel 2b) profiles for thermally-driven wind, where \(r_{\rm sb}\) and \(v_\infty\) are equal to those from table \ref{tab:param}. The red curve represents the~\citetalias{CC1985} model for which radiative cooling is turned off, whereas green and blue curves do account for radiative cooling. The wind represented by the blue curve cools significantly more rapidly than that for the green curve, as both its mass and energy injection rate, \(\dot{M}\) and \(\dot{E}\), are twice as much.
}
\label{fig:cool}
\end{figure*}

Radiative cooling was not included in the outflow model of \citetalias{CC1985}. 
Cooling effects generally depend on both temperature and density, 
  and so could have substantial effects on the hydrodynamic structure 
  of a galactic wind.  
For instance, thermal free cooling is strongly dependent on density 
  and would significantly affect systems with high mass outflow rates.  
\citet[][hereafter~\citetalias{Silich2004}]{Silich2004} developed a semi-analytical model that takes radiative cooling into account by adding an extra cooling term to equation \ref{eq:CCenergy},
\begin{equation}
\label{eq:coolenergy}
\frac{1}{r^2}\frac{\rm d}{{\rm d}r}\left\{\rho vr^2\left(\frac{v^2}{2}+\frac{\gamma_{\rm g}}{\gamma_{\rm g}-1}\frac{P}{\rho}\right)\right\}=Q_{\rm th}-C \ , 
\end{equation}
where \(C=\rho^2\Lambda/\mu^2\) is the cooling rate, and $\mu = 1.4\,m_{\rm H}$ is the mean particle mass. \citetalias{Silich2004} adopted the cooling function \(\Lambda\left(T\right)\) from \cite{Raymond1976} (assuming solar metallicity).
Note that we use \textsc{Cloudy} \citep{Ferland2017} to generate a cooling function that covers a wider temperature range.
The non-linearity of the cooling function necessitates a numerical approach in order to solve the HD equations. 
We start by substituting equations \ref{eq:CCmass} and \ref{eq:CCmomentum} into equation \ref{eq:coolenergy},
\begin{equation}
\label{eq:cooldvdr}
\frac{{\rm d}v}{{\rm d}r}=\begin{dcases}
\frac{\left(\gamma_{\rm g}-1\right)\left(Q_{\rm th}-C\right)+q\left(\frac{\gamma_{\rm g}+1}{2}v^2-\frac{2}{3}{c_{\rm s}}^2\right)}{\rho\left({c_{\rm s}}^2-v^2\right)}&\text{if }r\leq r_{\rm sb},\\
\frac{\left(\gamma_{\rm g}-1\right)rC+2\gamma_{\rm g}vP}{r\rho\left(v^2-{c_{\rm s}}^2\right)}&\text{if }r>r_{\rm sb}.
\end{dcases}
\end{equation}
From equation \ref{eq:CCmomentum}, the pressure gradient then follows as
\begin{equation}
\frac{{\rm d}P}{{\rm d}r}=\begin{dcases}
-qv-\rho v\frac{{\rm d}v}{{\rm d}r}&\text{if }r\leq r_{\rm sb} \ , \\
-\rho v\frac{{\rm d}v}{{\rm d}r}&\text{if }r>r_{\rm sb} \ ,
\end{dcases}
\label{eq:cooldPdr}
\end{equation}
and according to equation \ref{eq:CCmass}, the density can be expressed as
\begin{equation}
\rho=\begin{dcases}
\frac{qr}{3v}&\text{if }r\leq r_{\rm sb} \ , \\
\frac{q{r_{\rm sb}}^3}{3r^2v}&\text{if }r>r_{\rm sb} \ .
\end{dcases}
\label{eq:coolrho}
\end{equation}
These are solved by use of a Runge-Kutta method~\citep[e.g.][]{Press2007}, adopting the following boundary condition at $r=0$:
\begin{equation}
\label{eq:coolrho0}
\rho_0=\mu\sqrt{\frac{Q_{\rm th}-q{c_{\rm s, 0}}^2/\left(\gamma_{\rm g}-1\right)}{\Lambda\left(T_0\right)}} \ , 
\end{equation}
which is obtained by taking the limits of equations \ref{eq:cooldvdr} and \ref{eq:coolrho} as \(r,v\to0\). 
Using equation \ref{eq:coolrho0}, equation \ref{eq:CCcs} and the ideal gas law, the initial conditions \(\rho_0\), \(c_{\rm s,0}\) and \(P_0\) can be determined by using the appropriate \(T_0\). We note that the sign of the denominator in equation \ref{eq:cooldvdr} changes across \(r_{\rm sb}\), which implies that \(v=c_{\rm s}\) has to hold at \(r_{\rm sb}\) (or equivalently \(r_{\rm s}=r_{\rm sb}\)) for a physical outflow solution. The true boundary condition can then be found by iterating \(T_0\) until \(r_{\rm s}\) converges to \(r_{\rm sb}\). The process of finding the correct boundary condition in this way is detailed further in~\citetalias{Silich2004}.

We solve the HD equations numerically, and show the resulting velocity and temperature profiles in Fig.~\ref{fig:cool} (panel 2a and 2b, respectively). Note that radiative cooling is turned off for the result shown by the red curve, thus giving a result equivalent to~\citetalias{CC1985}. Such an outflow wind expands adiabatically (\(P\propto\rho^{\gamma_{\rm g}}\)), and the temperature drops according to \(T\propto r^{-2\left(\gamma_{\rm g}-1\right)}\). When the radiative cooling is present, the wind temperature \(T\left(r\right)\) cools more rapidly when the mass injection rate is higher. This is because the cooling rate \(C=\rho^2\Lambda/\mu^2\) is scaled in proportion to the mass injection rate, in line with equation \ref{eq:coolrho}. We see this effect in panel 2b, where the blue curve falls much more rapidly than the green curve (for which the SFR is twice as much). Additionally, the mass injection rate has an upper limit, beyond which there will be cooling instability within the outflow. Such limit scales as $\dot{M}_{\rm max}\propto r_{\rm sb}$.

\subsection{Radiation-driven outflows}


\subsubsection{Dusty shells}
\label{sec:Thompson}
\citealt{Thompson2015} (hereafter~\citetalias{Thompson2015})
considered the momentum transfer arising from a point-like source of radiation into dusty clouds to drive an outflow (in lieu of the thermal pressure gradient invoked formerly in~\citetalias{CC1985}). The momentum equation is written as
\begin{equation}\label{eq:Thompson}
\rho v\frac{{\rm d}v}{{\rm d}r}=\rho f_{\rm rad}+\rho f_{\rm grav}=\frac{\rho\kappa L}{4\pi r^2c}-\frac{GM\rho}{r^2} \  
\end{equation}
(cf. equation \ref{eq:CCmomentum}). 
The driving force delivered by the radiation \(\rho f_{\rm rad}\) is governed by the density of the outflow wind material, $\rho$, the radiation energy density, $L/4\pi r^2 c$ (where ${c}$ is the speed of light), and the interaction cross section between the dust-enriched wind and the radiation, as characterised by the mean opacity over all wavelengths, $\kappa$.

\begin{figure}
\includegraphics[width=\columnwidth]{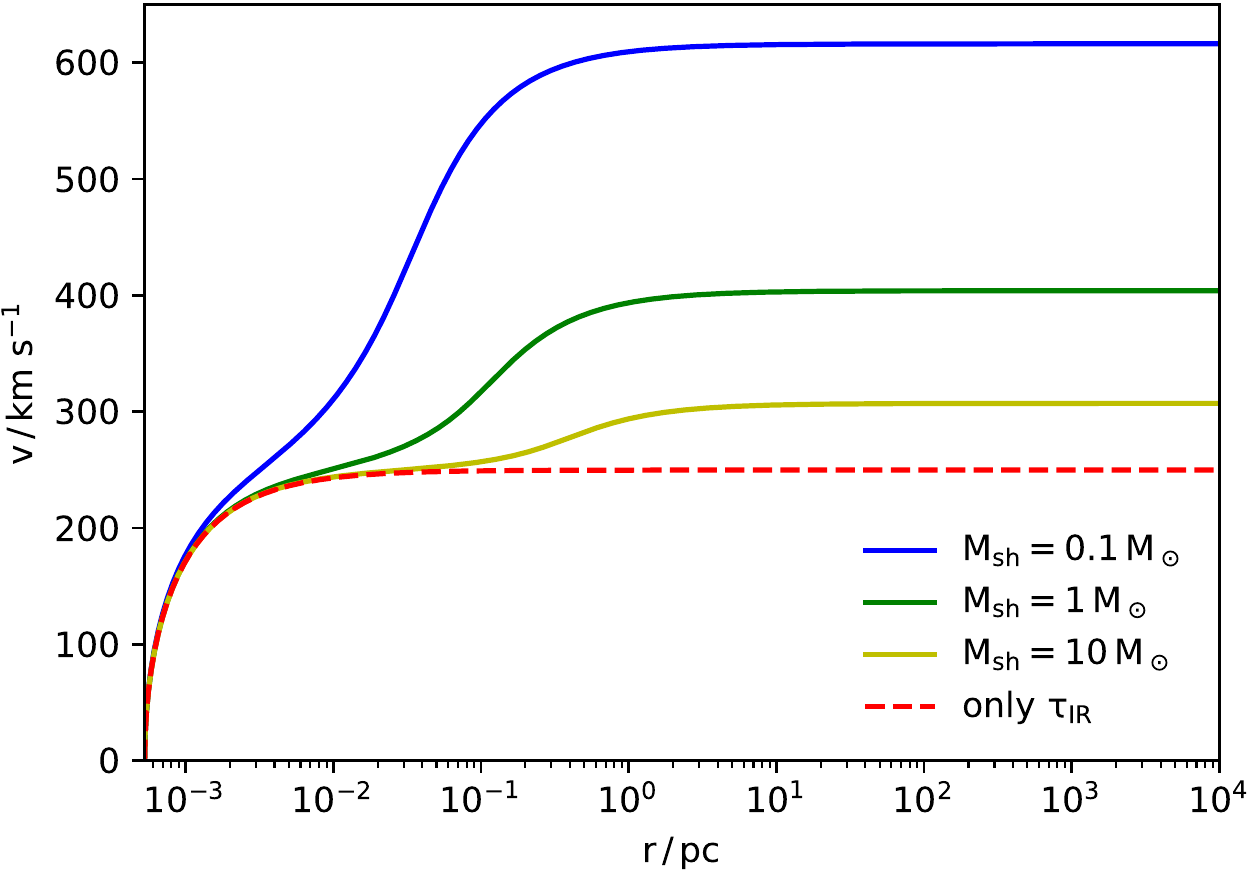}
\caption{Velocity profile of dusty shells of mass 0.1, 1 and 10\(\,\rm M_\odot\) driven by stellar radiation from a massive star with \(M=100\,\rm M_\odot\) and \(L=10^7\,\rm L_\odot\), where \(r_0=r_{\rm sub}\), \(\tau_{\rm IR}\left(r_0\right)=300\times M_{\rm sh}\) and \(\tau_{\rm UV}=250\,\tau_{\rm IR}\). The effect of IR contribution is shown by the dashed curve, while the solid curves account also for the UV contribution. We note that the shell velocity is inversely correlated to \(M_{\rm sh}\).
}
\label{fig:Thompson}
\end{figure}

\citetalias{Thompson2015} specifically modelled a dusty shell of mass \(M_{\rm sh}\) driven by radiation from a massive star (\(M=100\,\rm M_\odot\), \(L_{\rm UV}=10^7\,\rm L_\odot\)). 
The dust shell initially develops at a distance \(r_0\) from the star, being determined by the the dust sublimation radius \(r_{\rm sub}\), i.e.
\begin{align}
r_0 &= r_{\rm sub} \nonumber \\
 & = 5.28\times 10^{-4} \;\! \left(\frac{L_{\rm UV}}{10^7\;\!\text{L}_{\odot}} \right)^{1/2} \;\! \left( \frac{T_{\rm sub}}{1,500\,\rm K} \right)^{-2}\,\rm pc \ ,
\end{align}
with the dust component being evaporated at closer distances.
For a sublimation temperature of 
\(T_{\rm sub}=1500\,\)K, the initial distance is \(r_0=5.28\times10^{-4}\,\)pc.
The stellar spectrum peaks in the ultra-violet (UV) band, however much of this radiation is reprocessed to infra-rad (IR) by the dust. The effects of the incident UV and reprocessed IR contributions may be quantified together as  
\begin{equation} 
\label{eq:UVIR} 
 M_{\rm sh}v
 \frac{{\rm d}v}{{\rm d}r}
=\left(1-e^{-\tau_{\rm UV}} +\tau_{\rm IR}\right)
\frac{L_{\rm UV}}{c} \ , 
\end{equation}  
with the terms on the right hand side accounting for the UV irradiation and its attenuation, and the re-radiated IR respectively.
The UV and IR optical depths in the shell are given by
\begin{equation}
\tau_{\rm i}=\frac{\kappa_{\rm i}M_{\rm sh}}{4\pi r^2}=\frac{{r_{\rm i}}^2}{r^2} \ ,
\end{equation}
with subscript ${\rm i}$ denoting either the UV or IR contribution as required.
At the initial (sublimation) radius \(r_0\), \citetalias{Thompson2015} adopts optical depths of 
\(\tau_{\rm IR}\left(r_0\right)=300\times M_{\rm sh}\) and \(\tau_{\rm UV}=250\,\tau_{\rm IR}\) (which initially yields an optically thick shell to both UV and IR, which becomes optically thin at larger distances), and considers shell masses of 0.1, 1 and 10 \(\rm M_\odot\) to solve equation~\ref{eq:UVIR} subject to the boundary condition that $v\left(r_0\right) = 0$. This gives
\begin{equation}\label{eq:UVIRsolve}
v^2\left(r\right)=\frac{2rL_{\rm UV}}{M_{\rm sh}c}\left[H\left(r\right)-H\left(r_0\right)\right]-\frac{\kappa_{\rm IR}L_{\rm UV}}{2\pi c}\left(\frac{1}{r}-\frac{1}{r_0}\right) \ ,
\end{equation}
where \(H\left(r\right)\) is a scale factor as a function of \(r\), derived from solving the UV irradiation and its attenuation, given by
\begin{equation}
H\left(r\right)=1-\exp\left(-\frac{{r_{\rm UV}}^2}{r^2}\right)-\frac{\sqrt{\pi}r_{\rm UV}}{r}{\rm erf}\left(\frac{r_{\rm UV}}{r}\right) \ , 
\end{equation}
where erf is the error function. We plot the shell velocity according to the~\citetalias{Thompson2015} prescription in Fig.~\ref{fig:Thompson} where the acceleration is dominated by the IR contribution at small distances, with further UV acceleration arising after the shell becomes optically thin to IR radiation. A terminal velocity is attained after the shell also becomes thin to the UV contribution.
We note that the regions governed by the IR driving effect (denoted by the dashed red line in Fig.~\ref{fig:Thompson}) are independent of the shell mass, $M_{\rm sh}$. The wind profile is also independent of \(r_{\rm sb}\), because the radiation arises from a point-source instead of a starburst injection zone.

While \citetalias{Thompson2015} calculated the kinematics of the dusty shell, \citealt{Sharma2013} (hereafter~\citetalias{Sharma2013}) showed that the calculation can be generalised by modifying the HD equations \ref{eq:CCmass}, \ref{eq:CCmomentum} and \ref{eq:CCenergy} to account for the impact of the radiative driving. Under their prescription, the HD equations of~\citetalias{CC1985} are adopted and modified to 
\begin{gather}
\label{eq:radmomentum}
\rho v\frac{{\rm d}v}{{\rm d}r}=\rho f-\frac{{\rm d}P}{{\rm d}r}-qv \ , \\
\label{eq:radenergy}
\frac{1}{r^2}\frac{\rm d}{{\rm d}r}\left(\rho vr^2\left(\frac{v^2}{2}+\frac{\gamma_{\rm g}}{\gamma_{\rm g}-1}\frac{P}{\rho}\right)\right)=Q_{\rm th}+\rho Fv \ , 
\end{gather}
where the external force term is \(f=f_{\rm rad}+f_{\rm grav}\) and external power term is \(Fv=F_{\rm rad}v+f_{\rm grav}v\). \citetalias{Sharma2013} specifically considered radiation driven by an active galactic nucleus (AGN) so that \(f_{\rm rad}=F_{\rm rad}\), and solved the equations numerically by adopting the boundary condition that \(r_{\rm s}=r_{\rm sb}\). In the following, we derive from this our model with a starburst nucleus instead of an AGN.

\subsubsection{Galactic winds}
Equations \ref{eq:radmomentum} and \ref{eq:radenergy} can be solved to give
\begin{equation}
\frac{{\rm d}v}{{\rm d}r}=\begin{dcases}
\frac{\left(\gamma_{\rm g}-1\right)\left(Q_{\rm th}+\rho Fv\right)-\gamma_{\rm g}\rho fv+q(\frac{\gamma_{\rm g}+1}{2}v^2-\frac{2}{3}{c_{\rm s}}^2)}{\rho\left({c_{\rm s}}^2-v^2\right)} \ , \\
\frac{r\gamma_{\rm g}\rho fv-r\left(\gamma_{\rm g}-1\right)\rho Fv+2\gamma_{\rm g}vP}{r\rho\left(v^2-{c_{\rm s}}^2\right)} \ ,
\end{dcases}
\label{eq:raddvdr}
\end{equation} 
for $r\leq r_{\rm sb}$ and $r>r_{\rm sb}$ respectively, and
\begin{equation}
\label{eq:raddPdr}
\frac{{\rm d}P}{{\rm d}r}=\begin{dcases}
\rho f-qv-\rho v\frac{{\rm d}v}{{\rm d}r}&\text{if }r\leq r_{\rm sb} \ , \\
\rho f-\rho v\frac{{\rm d}v}{{\rm d}r}&\text{if }r>r_{\rm sb} \ ,
\end{dcases}\end{equation}
which may be compared to equations \ref{eq:cooldvdr} and \ref{eq:cooldPdr} in \citetalias{Silich2004}, where radiative forces are absent. 
The inclusion of radiative forces complicates the solution, because the numerator in equation \ref{eq:raddvdr} can now become negative at $r<r_{\rm sb}$ 
if the energy density of the radiation is larger than 
 the thermal energy density of the fluid. 
This causes the sonic radius to fall within the starburst radius ($r_{\rm s}<r_{\rm sb}$). 
The HD variables $v, \rho, P$ and ${\rm d}v/{\rm d}r$ can be calculated given the value of $r_{\rm s}$ as both the numerator and denominator of equation \ref{eq:raddvdr} vanish at the limit \(r\to r_{\rm s}\). This allow $r_{\rm s}$ to be determined by iteration until a solution is found where $v\rightarrow0$ when $r\rightarrow0$.
This approach is similar to that adopted in~\citet{Silich2011}.
If there is sufficient thermal energy injection to enable \(r_{\rm s}=r_{\rm sb}\),  the correct \(\rho_0\) at \(r=0\) may be found iteratively by changing \(\rho_0\) until \(r_{\rm s}\) converges to \(r_{\rm sb}\).\footnote{Note that from equation \ref{eq:coolrho}, \({\rm d}v/{\rm d}r\) at \(r=0\) is equal to \(q/(3\rho_0)\).} 


 
\label{sec:rad}
\begin{figure}
\includegraphics[width=\columnwidth]{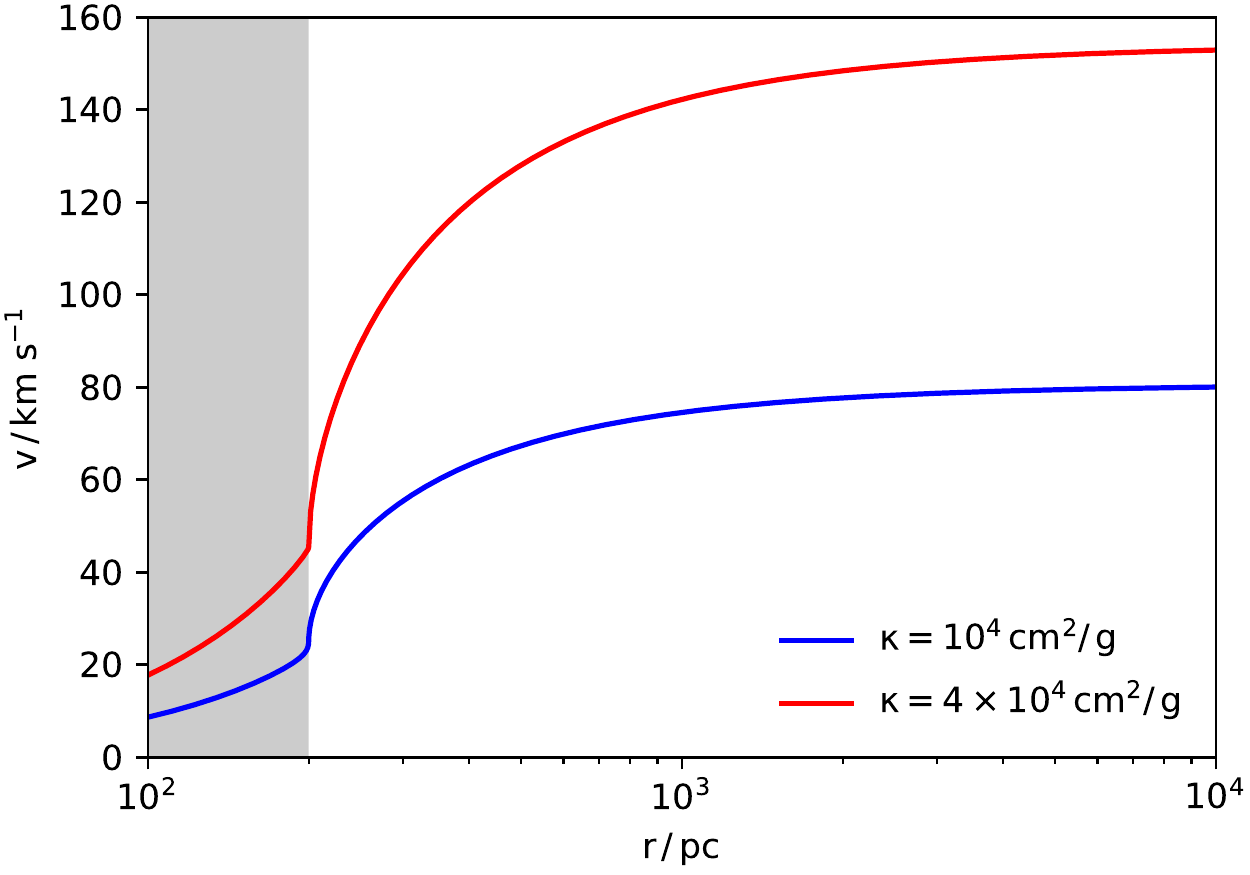}
\caption{Velocity profile of a radiation-driven wind using parameters from Table~\ref{tab:param}, where the energy injection is radiation dominated (\(\dot{E}_{\rm rad}=99\%\times\dot{E}\)). The mean opacity is \(\kappa=10^4\,\rm cm^2/g\) for the blue curve, and four times higher for the red curve. The wind speed in the red curve is approximately twice that of the blue curve, consistent with the scaling relation \(u^2\propto\kappa\). Note that the velocity gradient is discontinuous across \(r_{\rm sb}\).}
\label{fig:radvelocity}
\end{figure}
 
 Following equation~\ref{eq:Thompson}, the radiative force due to a starburst core in a galaxy may be expressed as
 \begin{equation}\label{eq:frad}
f_{\rm rad}=\begin{dcases}
\frac{\kappa Lr}{4\pi{r_{\rm sb}}^3c} ,&\text{if }r\leq r_{\rm sb} \ , \\
\frac{\kappa L}{4\pi r^2c},&\text{if }r>r_{\rm sb} \ , 
\end{dcases}\end{equation} 
 which is similar to the approach taken by \citetalias{Sharma2013}.
Some of the energy from the radiation from the core is thermalised directly into the wind to contribute to the thermal gas pressure gradient, while some transfers momentum directly.
The effective radiative power is
\begin{equation}
\label{eq:Frad}
F_{\rm rad}v = \frac{3\kappa Lv}{16\pi {r_{\rm sb}}^3cr}\left(2r_{\rm sb}r-\left(r^2-{r_{\rm sb}}^2\right)\ln{\frac{r+r_{\rm sb}}{|r-r_{\rm sb}|}}\right) \ , 
\end{equation} 
where \(F_{\rm rad}=f_{\rm rad}\) at the limit where \(r_{\rm sb}\to 0\).    
We parametrise $L$, the luminosity of the radiation generated, in terms of the energy injection rate $\dot{E}$ and the thermal, mechanical outflow of the fluid, i.e.,
\begin{equation}
\label{eq:radL}
L=\dot{E}-\dot{M}\left(\frac{v^2}{2}+\frac{{c_{\rm s}}^2}{\gamma_{\rm g}-1}\right)\ . 
\end{equation}
We further parametrise the relative distribution of energy injected into radiation and the thermal fluid component, which are represented by two terms, with $\dot{E}_{\rm rad}+\dot{E}_{\rm th} = {\dot E}$.

Gravitational effects due to dark matter (DM) can be considered by adopting an NFW \citep{Navarro1996} density profile, such that the associated force may be written as
\begin{gather}
\label{eq:fgrav}
f_{\rm grav}=-\frac{GM\left(r\right)}{r^2}=-\frac{GM_{\rm tot}\xi\left(r\right)}{r^2\xi\left(R_{\rm vir}\right)} \ , \\
\xi\left(r\right)=\ln\frac{R_{\rm s}+r}{R_{\rm s}}-\frac{r}{R_{\rm s}+r} \ , 
\end{gather}
where \(R_{\rm vir}\) is the virial radius, \(M_{\rm tot}=M\left(R_{\rm vir}\right)\) is the total halo mass, and \(R_{\rm s}\) is its scale radius.

We show the solution in Fig.~\ref{fig:radvelocity} for the radiation-driven outflow, adopting standard parameter values for \(r_{\rm sb}\), \(\dot{M}\), \(\dot{E}\), \(M_{\rm tot}\), \(R_{\rm s}\) and \(R_{\rm vir}\) (see table \ref{tab:param}), and where the injection of energy is dominated by radiation (\(\dot{E}_{\rm rad}=99\%\times\dot{E}\)). Since thermal energy is negligible, the wind reaches a supersonic velocity below \(r_{\rm sb}\). This causes the velocity gradient to be finite at \(r_{\rm sb}\) (contrary to the case in section \ref{sec:thermal}) and behave as a step function across \(r_{\rm sb}\). Note that \({u_{\infty}}^2\leq 2\dot{E}/\dot{M}\), as the radiation can escape the galaxy without getting absorbed. If the wind is optically thin to the radiation, it can be shown from equation \ref{eq:radmomentum} that \(u^2\propto\kappa\).

\subsection{Cosmic ray outflows}


\subsubsection{HD model}
\label{sec:CR}

Around 10\% of the energy of a SN event is passed
to cosmic rays\footnote{Variation of this percentage is seen in both theoretical and observational works, suggesting a range between 
7\%~\citep{Lemoine2012A&A} and 30\%~\citep{Caprioli2012JCAP, Fields2001A&A}. 10\% is usually suggested or taken as a characteristic value~\citep[e.g.][]{Helder2009, Dermer2013A&A, Morlino2012A&A, Strong2010ApJ, Wang2018MNRAS}.} (CRs) which can contribute towards driving an outflow.
\citealt{Ipavich1975} (hereafter~\citetalias{Ipavich1975}) considered the injection of CRs which couple to the ionised wind fluid. The momentum equation is written as
\begin{equation}
\label{eq:CRmomentum}
\rho v\frac{{\rm d}v}{{\rm d}r} = -\frac{{\rm d}P}{{\rm d}r}-\frac{{\rm d}P_{\rm C}}{{\rm d}r}+\rho f_{\rm grav} \ . 
\end{equation}

The CRs form a relativistic non-thermal component of the wind, which have negligible bulk kinetic energy but non-negligible pressure. Energy is transferred from the relativistic component to the thermal component of the wind at a rate of \(I=-\left(v+v_{\rm A}\right)\cdot \nabla P_{\rm C}\) (where $v$ is the bulk flow velocity and $v_{\rm A}$ is the Alfv\'en velocity), allowing the energy equations to be written as 
\begin{gather}
\label{eq:CRenergy}
\frac{1}{r^2}\frac{\rm d}{{\rm d}r}\left(\rho vr^2\left(\frac{v^2}{2}+\frac{\gamma_{\rm g}}{\gamma_{\rm g}-1}\frac{P}{\rho}\right)\right)=I+\rho f_{\rm grav}v \ , \\
\label{eq:CRmagenergy}
\frac{1}{r^2}\frac{\rm d}{{\rm d}r}\left(\rho \left(v+v_{\rm A}\right)r^2\frac{\gamma_{\rm C}}{\gamma_{\rm C}-1}\frac{P_{\rm C}}{\rho}\right)=-I \ , \\
\label{eq:CRtransfer}
I=-\left(v+v_{\rm A}\right)\frac{{\rm d}P_{\rm C}}{{\rm d}r} \ ,  
\end{gather}
where \(\gamma_{\rm C}=4/3\) is the relativistic adiabatic index and $v_{\rm A}$ is the local Alfv\'en speed.  
We consider that the magnetic field within the flow 
  is highly tangled. 
Thus, the Alfv\'en speed takes an effective value,  
  i.e. setting $v_{\rm A} = {\tilde v}_{\rm A} = \sqrt{\langle B^2 \rangle}/\sqrt{4\pi \rho}$. 
Suppose that non-directional magnetic flux conservation holds 
  despite the tangled magnetic field structure, then we have 
\begin{equation}\label{eq:CRalfven}
{\tilde v}_{\rm A} =
 \frac{{r_0}^2\sqrt{\langle {B_0}^2\rangle}}{r^2\sqrt{4\pi\rho}}
  \ .  
\end{equation} 
In our formulation, magneto-hydrodynamic effects are ignored. 
The magnetic field simply acts as a mediator 
 between the CR and gas components, 
 through which energy transfer is facilitated. 
Thus, the exact local structure of the magnetic field is irrelevant.  
We may therefore adopt a parametrisation, 
  in which the field-gas coupling 
  depends only the local root mean square value of the magnetic field  
  (and hence the local averaged value of the magnetic energy density).

The effective sound speed \(c_*\) follows as \citep[see][]{Samui2010}
\begin{equation}
\label{eq:CRcs}
{c_*}^2=\frac{\gamma_{\rm g}P}{\rho}+\frac{\gamma_{\rm C}P_{\rm C}}{\rho}\frac{\left(2v+v_{\rm A}\right)\left(v-\left(\gamma_{\rm g}-1\right)v_{\rm A}\right)}{2v\left(v+v_{\rm A}\right)} \ , 
\end{equation}
which then yields \citep[following][for the derivation]{Samui2010}
\begin{gather}
\label{eq:CRdPcdr}
\frac{{\rm d}P_{\rm C}}{{\rm d}r}=\frac{\gamma_{\rm C}P_{\rm C}}{\rho}\frac{2v+v_{\rm A}}{2\left(v+v_{\rm A}\right)}\frac{{\rm d}\rho}{{\rm d}r} \ , \\
\label{eq:CRdPdr}
\frac{{\rm d}P}{{\rm d}r}=\left(\frac{\gamma_{\rm g}P}{\rho}-\frac{\gamma_{\rm C}P_c}{\rho}\frac{\gamma_{\rm g}-1}{2}\frac{v_{\rm A}\left(2v+v_{\rm A}\right)}{v\left(v+v_{\rm A}\right)}\right)\frac{{\rm d}\rho}{{\rm d}r} \ , \\
\label{eq:CRdvdr}
\frac{{\rm d}u}{{\rm d}r}=\frac{2v{c_*}^2+rf_{\rm grav}v}{r\left(v^2-{c_*}^2\right)} \ .
\end{gather}

\begin{table*}
\centering
\renewcommand\theadset{\def\arraystretch{0.9}}
\begin{tabular}{*{7}{c}}
Section & Model / Reference & \thead{Driving \\ mechanism} & \thead{Convergent \textsuperscript{\textit{a}} \\ at \(r=0\)} & \thead{Outflow \textsuperscript{\textit{b}} \\ temperature} & \thead{Radiative \textsuperscript{\textit{c}} \\ cooling} & \thead{Gravitational \\ potential} \\
\midrule
\ref{sec:CC85} & \cite{CC1985} & Thermal pressure & \ding{51} & \ding{51} & \ding{55} & \ding{55} \\
\ref{sec:radcool} & \cite{Silich2004} & Thermal pressure & \ding{51} & \ding{51} & \ding{51} & \ding{55} \\
\ref{sec:Thompson} & \cite{Thompson2015} & Radiative pressure & \ding{55} & \ding{55} & \ding{55} & \ding{51} \\
\ref{sec:Thompson} & \cite{Sharma2013} & Radiative pressure & \ding{55} & \ding{51} & \ding{55} & \ding{51} \\
\ref{sec:rad} & Yu et al. (2019) & Radiative pressure & \ding{51} & \ding{51} & \ding{55} & \ding{51} \\
\ref{sec:CR} & \cite{Ipavich1975}, \cite{Samui2010} & Cosmic rays & \ding{55} & \ding{51} & \ding{55} & \ding{51} \\
\ref{sec:CRBC} & Yu et al. (2019) & Cosmic rays & \ding{51} & \ding{51} & \ding{55} & \ding{51} \\
\ref{sec:mix} & Yu et al. (2019) & Combined & \ding{51} & \ding{51} & \ding{51} & \ding{51} \\
\end{tabular}
\caption{An outline of the HD models (including both the original models and the modifications adopted) discussed in section~\ref{sec:model}. Notes: \newline
\textsuperscript{\textit{a}} For HD models which do not "converge at \(r=0\)", the calculated stationary solutions do not produce physical results at \(r=0\) (which may be justified if we are only concerned with the outflow profiles at larger scales). Although there are infinitely many initial conditions to choose from, continuity in theory requires that $u=0$ at $r=0$. The other boundary condition at $r_{\rm sb}$ is a natural choice, as $q_{\rm m}$ and $q_{\rm e}$ change abruptly across the boundary of the starburst injection zone.
\newline
\textsuperscript{\textit{b}} The dusty wind model in \protect\cite{Thompson2015} is a kinematic model that does not take account of the temperature of the fluid. \newline
\textsuperscript{\textit{c}} It is possible to include radiative cooling as part of the modifications we introduced, but the stationary solution of the HD equations is inhibited if the wind is not hot enough, due to cooling instability (a known issue as discussed in \protect\citealt{Silich2004}). We therefore only turn radiative cooling on if the corresponding stationary solution exists, which will be discussed in greater detail in section \ref{sec:remark}.}
\label{tab:model}
\end{table*}

\citetalias{Ipavich1975} solved these by establishing a critical point at which both the numerator and denominator of equation \ref{eq:CRdvdr} vanish. This leads to \(2v^2=2{c_*}^2={V_{\rm c}}^2\), where \(V_{\rm c}=\sqrt{GM/r}\) is the circular velocity, and this defines both \(v\) and \(c_*\) at this critical point. However, the resulting solution is divergent towards \(r=0\) and it requires the gravitational force to be strong, contrary to \citetalias{CC1985}. We show in the following a derivation of the boundary condition in the case where \(f_{\rm grav}\) is small.

\subsubsection{Boundary condition}
\label{sec:CRBC}
\begin{figure}
\includegraphics[width=\columnwidth]{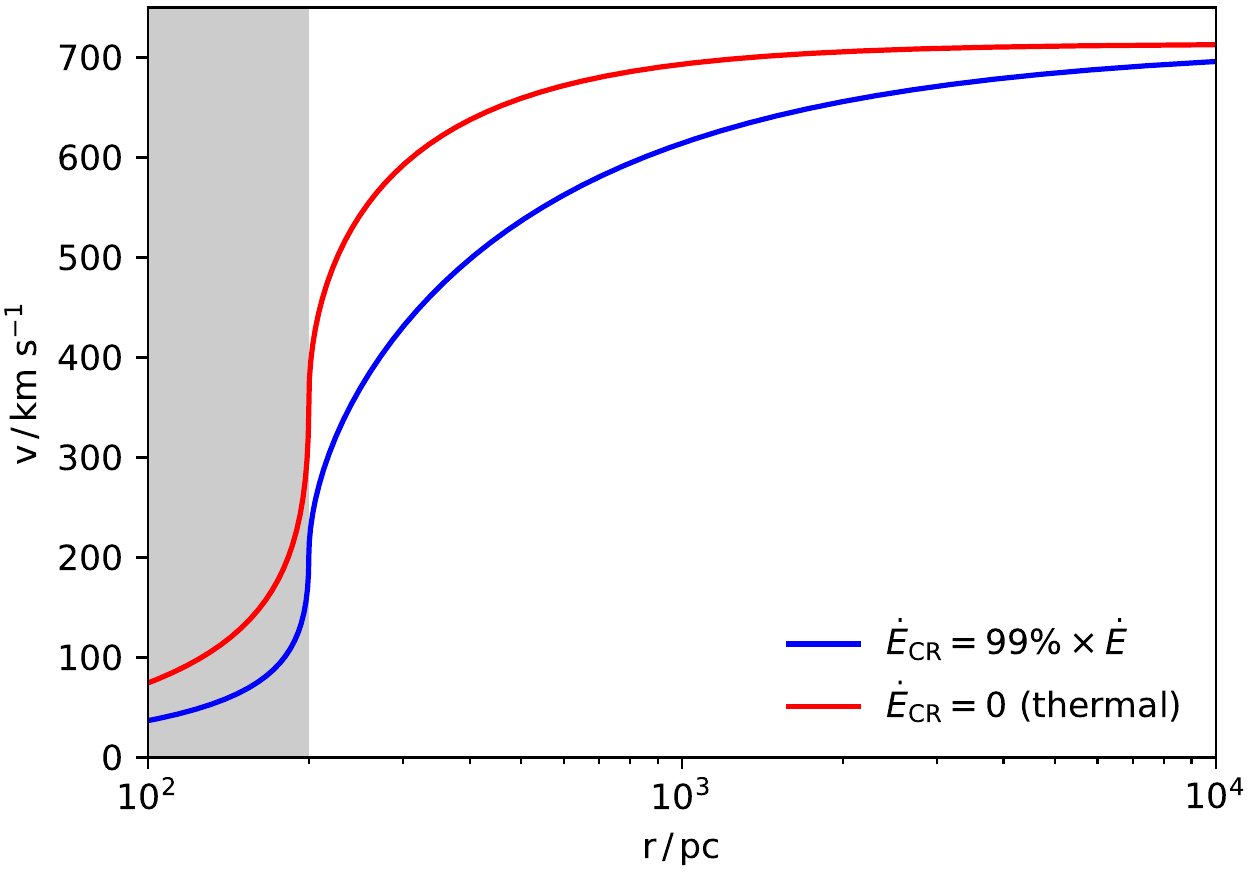}
\caption{Velocity profile of the CR-driven wind (blue, \(\dot{E}_{\rm CR}\approx\dot{E}\)) and thermally-driven wind (red, \(\dot{E}_{\rm th}=\dot{E}\)), using standard parameter choices (see Table~\ref{tab:param}). Note that \(\gamma_{\rm C}<\gamma_{\rm g}\), \(P_{\rm C}\) falls at a slower rate compared to \(P\) with galactocentric distance. As a result, the cosmic rays drive the outflow less effectively than the thermal gas pressure, as seen in the plot.}
\label{fig:CRvelocity}
\end{figure}

As in section \ref{sec:thermal}, we consider a starburst nucleus and establish the boundary condition \(v=c_*\) at \(r_{\rm sb}\). The HD equations for \(r\leq r_{\rm sb}\) are
\begin{gather}
\label{eq:inCRdPcdr}
\frac{{\rm d}P_{\rm C}}{{\rm d}r}=\frac{\gamma_{\rm C}P_{\rm C}}{\rho}\frac{2v+v_{\rm A}}{2\left(v+v_{\rm A}\right)}\frac{{\rm d}\rho}{{\rm d}r}+C_2 \ , \\
\label{eq:inCRdPdr}
\frac{{\rm d}P}{{\rm d}r}=\left(\frac{\gamma_{\rm g}P}{\rho}-\frac{\gamma_{\rm C}P_{\rm C}}{\rho}\frac{\gamma_{\rm g}-1}{2}\frac{v_{\rm A}\left(2v+v_{\rm A}\right)}{v\left(v+v_{\rm A}\right)}\right)\frac{{\rm d}\rho}{{\rm d}r}+C_1 \ , \\
\label{eq:inCRdvdr}
\frac{{\rm d}v}{{\rm d}r}=\frac{v{c_*}^2-rf_{\rm grav}v}{r\left({c_*}^2-v^2\right)}+\frac{v\left(C_1+C_2+qv\right)}{\rho\left({c_*}^2-v^2\right)} \ , 
\end{gather}
with \(C_1\) and \(C_2\) defined as
\begin{gather}
C_1=-\frac{3\gamma_{\rm g}P}{r}+\left(\gamma_{\rm g}-1\right)\left(\frac{Q_{\rm th}}{v}+\frac{3\rho v^2}{2r}-\frac{v_{\rm A}}{v}C_2\right) \ , \\
C_2=-\frac{3\gamma_{\rm C}P_{\rm C}}{r}+\frac{\left(\gamma_{\rm C}-1\right)Q_{\rm CR}}{v+v_{\rm A}} \ . 
\end{gather}
We construct a boundary condition by starting a numerical integration at \(r=0\).
\footnote{There are many valid approaches. We use a characteristic magnetic field strength following \(\langle |B| \rangle={B_0 \,r}/r_{\rm sb}\) at \(r\leq r_{\rm sb}\) to allow us to find a boundary condition in a tractable manner. We note that this approach is not physical -- indeed,
we would expect a roughly uniform-strength characteristic mean magnetic field to develop
within the starburst radius. However, this region is not of interest to our current work:
we do not require the description of the magnetic field inside the starburst region to be realistic, as the external HD results are unaffected by the method adopted to set a boundary condition.} 
By taking the limit $r\rightarrow 0$, equations \ref{eq:CRenergy}, \ref{eq:CRmagenergy} and \ref{eq:inCRdvdr} can be solved to give
\begin{gather}
\label{eq:CRinitPc}
\frac{\gamma_{\rm C}}{\gamma_{\rm C}-1}\frac{v+v_{\rm A}}{v}\frac{P_{\rm C,0}}{\rho_0}=\frac{Q_{\rm CR}}{q} \ , \\
\label{eq:CRinitP}
\frac{\gamma_{\rm g}}{\gamma_{\rm g}-1}\frac{P_0}{\rho_0}=\frac{Q_{\rm th}}{q} \ , 
\end{gather}
where both \(P_{\rm C,0}\) and \(P_0\) (i.e. their values at \(r=0\)) must be found by iteratively varying \(\rho_0\) until a condition is reached such that a solution arises where \(r_{\rm s}\) converges to \(r_{\rm sb}\). 
We note that \(P_{\rm C}\) falls at a slower rate with $r$ than \(P\) (because \(\gamma_{\rm C}<\gamma_{\rm g}\)). This means that the CR component is less effective in driving the wind compared to the thermal gas -- this is seen in Fig.~\ref{fig:CRvelocity}, where we show that the velocity profile for the CR-driven outflow is lower than the thermally-driven case at all altitudes.

\subsection{Generalised outflow model}
\label{sec:mix}
Thermal energy, radiation and CRs can simultaneously drive outflows. In this section, we introduce a general model which accounts for all of these simultaneously. Just as~\citetalias{Sharma2013} included radiative driving in the~\citetalias{CC1985} thermally-driven HD model (see section \ref{sec:rad}), we include radiative driving in the CR-driven model (in section~\ref{sec:CR}). Moreover, radiative cooling (Section \ref{sec:radcool}) can be included into this generalised approach. The full set of HD equations is
\begin{gather}
\tag{\ref{eq:CCmass}}
\frac{1}{r^2}\frac{\rm d}{{\rm d}r}\left(\rho vr^2\right)=q \ , \\
\label{eq:mixmomentum}
\rho v\frac{{\rm d}v}{{\rm d}r}=-\frac{{\rm d}P}{{\rm d}r}-\frac{{\rm d}P_{\rm C}}{{\rm d}r}+\rho f_{\rm rad}+\rho f_{\rm grav}-qv \ , \\
\begin{split}
\label{eq:mixenergy}
\frac{1}{r^2}\frac{\rm d}{{\rm d}r}\left\{\rho vr^2\left(\frac{v^2}{2}+\frac{\gamma_{\rm g}}{\gamma_{\rm g}-1} \frac{P}{\rho}\right)\right\}& \\
=Q_{\rm th}&-C+\rho F_{\rm rad}v+\rho f_{\rm grav}v+I 
\ ,
\end{split} \\
\label{eq:mixmagenergy}
\frac{1}{r^2}\frac{\rm d}{{\rm d}r}\left(\rho \left(v+v_{\rm A}\right)r^2\frac{\gamma_{\rm C}}{\gamma_{\rm C}-1}\frac{P_{\rm C}}{\rho}\right)=Q_{\rm CR}-I \ , 
\end{gather}
where, for completeness, we have replicated earlier equations when they remain unchanged.  
Here, \(f_{\rm rad}\), \(F_{\rm rad}\), \(f_{\rm grav}\), and \(I\) can be found in equations \ref{eq:frad}, \ref{eq:Frad}, \ref{eq:fgrav}, and \ref{eq:CRtransfer} respectively. With CRs, the luminosity \(L\) from equation \ref{eq:radL} is modified to
\begin{equation}
\label{eq:mixL}
L=\dot{E}-\dot{M}\left(\frac{v^2}{2}+\frac{\gamma_{\rm g}}{\gamma_{\rm g}-1}\frac{P}{\rho}+\frac{\gamma_{\rm C}}{\gamma_{\rm C}-1}\frac{v+v_{\rm A}}{v}\frac{P_{\rm C}}{\rho}\right) \ , 
\end{equation}
where \(L\) is assumed to be unaffected by the gravitational potential and cooling effect. The resulting HD equations may be written as
\begin{gather}
\frac{{\rm d}P_{\rm C}}{{\rm d}r}=\frac{{\rm d}P_{\rm C}}{{\rm d}r}\biggl\rvert_{\rm CR} \ , \\
\frac{{\rm d}P}{{\rm d}r}=\frac{{\rm d}P}{{\rm d}r}\biggl\rvert_{\rm CR}+\left(\gamma_{\rm g}-1\right)\left(-\frac{C}{v}+\rho F_{\rm rad}-\rho f_{\rm rad}\right) \ , \\
\frac{{\rm d}v}{{\rm d}r}=\frac{{\rm d}v}{{\rm d}r}\biggl\rvert_{\rm CR}+\frac{\left(\gamma_{\rm g}-1\right)\left(-C+\rho F_{\rm rad}v\right)-\gamma_{\rm g}\rho f_{\rm rad}v}{\rho\left({c_*}^2-v^2\right)} \ , 
\end{gather}
where \(\frac{{\rm d}P}{{\rm d}r}\big\rvert_{\rm CR}\), \(\frac{{\rm d}P_{\rm C}}{{\rm d}r}\big\rvert_{\rm CR}\) and \(\frac{{\rm d}v}{{\rm d}r}\big\rvert_{\rm CR}\) are defined in equations \ref{eq:inCRdPdr}, \ref{eq:inCRdPcdr} and \ref{eq:inCRdvdr} respectively for \(r\leq r_{\rm sb}\) and \ref{eq:CRdPdr}, \ref{eq:CRdPcdr} and \ref{eq:CRdvdr} (respectively) for \(r>r_{\rm sb}\). 

\begin{figure*}
\includegraphics[width=2\columnwidth]{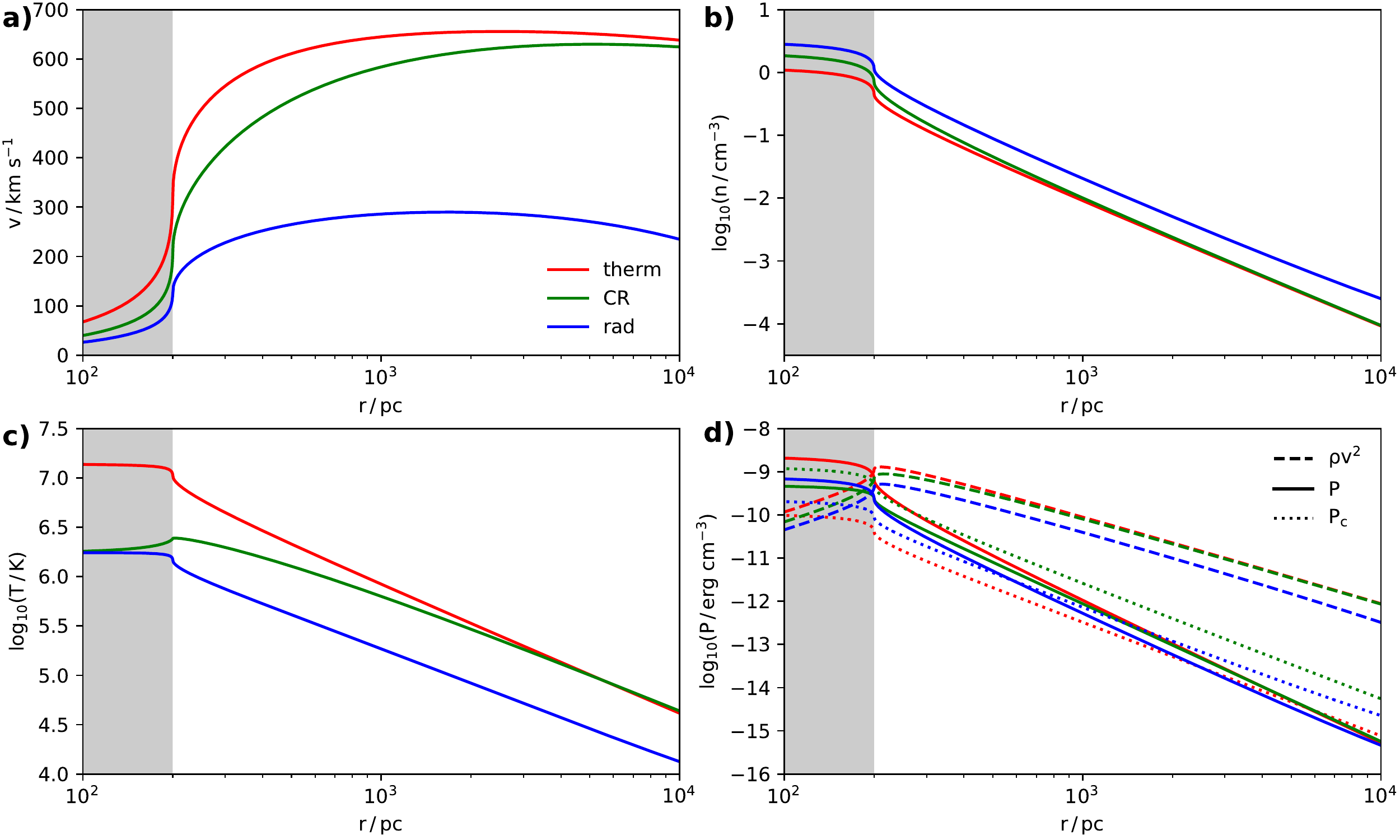}
\caption{Plots showing the \textbf{a)} velocity, \textbf{b)} density, \textbf{c)} temperature and \textbf{d)} pressure profiles of the galactic outflows for which the injection of energy is dominated by thermal energy (in red), radiation (in blue) and CRs (in green). In panel \textbf{d}, the dashed, solid and dotted lines represent the ram pressure, thermal pressure and CR pressure,  respectively. Note that the density, temperature and pressure are plotted in logarithmic scales. This shows that the thermally-driven wind is the hottest, the radiatively-driven wind is the coldest and most dense, and the CR-driven wind has the highest CR pressure. The standard parameters from Table~\ref{tab:param} were used.}
\label{fig:mechanism}
\end{figure*}

By the same process as described in section \ref{sec:CR}, we find the boundary conditions to be
\begin{gather}
\tag{\ref{eq:CRinitPc}}
\frac{\gamma_{\rm C}}{\gamma_{\rm C}-1}\frac{v+v_{\rm A}}{u}\frac{P_{\rm C,0}}{\rho_0}=\frac{Q_{\rm CR}}{q} \ , \\
\tag{\ref{eq:coolrho0}}
\frac{\gamma_{\rm g}}{\gamma_{\rm g}-1}\frac{P_0}{\rho_0}=\frac{Q_{\rm th}-C}{q} \ , 
\end{gather}
where \(\rho_0\) is determined again by iterating its values until a solution is found for which \(r_{\rm s}\) converges to \(r_{\rm sb}\). We plot and discuss the resulting solutions in section~\ref{sec:compare}.
A summary of the HD models discussed in this section is presented in table \ref{tab:model}.

\section{Results and discussion}
\label{sec:compare}

The generalised outflow model presented in section \ref{sec:mix} 
allows us to assess the role of each contributing driving mechanism. In practise, relevant parameters may be determined from the observed physical properties of outflows and their host galaxies. 
In this section, we demonstrate the effect of model parameters on the HD variables ($v$, $\rho$, $T$, and $P$) along an outflow.
The impact of different energy contributions to each of the wind components is considered 
 in section \ref{sec:mechanism}.
The effects of the galactic magnetic field strength ($B_0$) in CR-driven winds and
opacity ($\kappa$) in radiation-driven winds 
 are discussed in section \ref{sec:B_0} and \ref{sec:kappa} respectively, and gravitational effects (e.g. due to the dark matter halo) are explored in section \ref{sec:G}.
The values for the initial conditions can be found in Table~\ref{tab:init}.



\subsection{Driving mechanisms}
\label{sec:mechanism}

The reference parameters shown in table \ref{tab:param} are adopted as a baseline case, that takes into account the typical properties of nearby star-bursting galaxy M82. To simplify among the many possible scenarios, we compare the contribution to the outflows from thermal, radiation and CR physics by parameterising their respective energy injection rates, and explore three fiducial cases, namely 
\(\left(\dot{E}_{\rm th},\dot{E}_{\rm rad},\dot{E}_{\rm CR}\right)/\dot{E}=\left(0.8,0.1,0.1\right)\) for hot, thermally-dominated winds; \(\left(0.1, 0.8, 0.1\right)\) for radiation-driven winds; and \(\left(0.1,0.1,0.8\right)\) for CR-driven winds. 
The results are shown in Fig.~\ref{fig:mechanism} where the velocity, density, temperature and pressures profiles are plotted in panel \textbf{a}, \textbf{b}, \textbf{c} and \textbf{d} respectively.
These show that the driving effect of thermal pressure is more
effective in accelerating a wind. This is followed by CRs and then radiation.
The velocity profiles of the thermally-driven and CR-driven winds are consistent with the results shown in Fig.~\ref{fig:CRvelocity}, where the thermal gas pressure is most effective in driving an outflow due to the larger adiabatic index allowing for a higher expansion rate.
Radiation is the least effective driving mechanism, because the wind is optically thin to the radiation (if \(\kappa\) is not chosen to be unphysically large) so most of the radiative energy escapes the galaxy without contributing towards driving the wind. Radiation-driven winds are the most dense, followed by CR-driven then thermally-driven cases. The density profiles in Fig.~\ref{fig:mechanism}, panel \textbf{b} follow from the velocity profiles, given that \(\rho\propto u^{-1}\).

Fig.~\ref{fig:mechanism}, panel \textbf{c} shows that the thermally-driven wind is the hottest, with its temperature profile being almost identical to that of Fig.~\ref{fig:cool}. 
The temperature profiles of the thermally-driven and radiation-driven winds are similar, but the former is eight times hotter. This follows from the temperature profile being strongly influenced by \(\dot{E}_{\rm th}\). While \(\dot{E}_{\rm th}\) is the same for radiation-driven wind and CR-driven wind, the CR-driven system is hotter because of the transfer of energy from the CR fluid to the thermal gas (cf. equation \ref{eq:CRtransfer}). 
Such differences in the temperature profiles emerge more clearly at higher flow altitudes. This is because the transfer of energy is cumulative.
 Note that the effect of radiative heating can be seen from the slight increase of temperature on the blue curve at \(r<r_{\rm sb}\), which would otherwise be absent if \(F_{\rm rad}=f_{\rm rad}\) (compare equations \ref{eq:frad} and \ref{eq:Frad}).

Fig.~\ref{fig:mechanism}, panel \textbf{d} shows the effective outflow pressure contributions, and how this varies along the outflows. The results are consistent with the results in the other panels: the ram pressure \(P_{\rm ram}=\rho u^2\) is plotted with dashed lines and becomes dominant in regions where the flow velocity is greatest (cf. panels \textbf{a} and \textbf{b}); the thermal pressure \(P=\rho T/\gamma_{\rm g}\) is shown using solid lines, and is greatest at the base of the outflow where temperatures are highest (cf. Fig.~\ref{fig:mechanism}, panels \textbf{b} and \textbf{c}); the CR pressure \(P_{\rm C}\) is shown using dotted lines, and follows from equation \ref{eq:CRinitPc} and panel \textbf{b} (where \(P_{\rm C}/\rho\approx Q_{\rm CR}/q\)). Note that the decay rate of \(P_{\rm C}\) is lower than that of \(P\), which verifies the difference in their adiabatic expansion rate as follows from \(\gamma_{\rm C} < \gamma_{\rm g}\).

\subsection{Magnetic field}
\label{sec:B_0}

\begin{figure*}
\includegraphics[width=2\columnwidth]{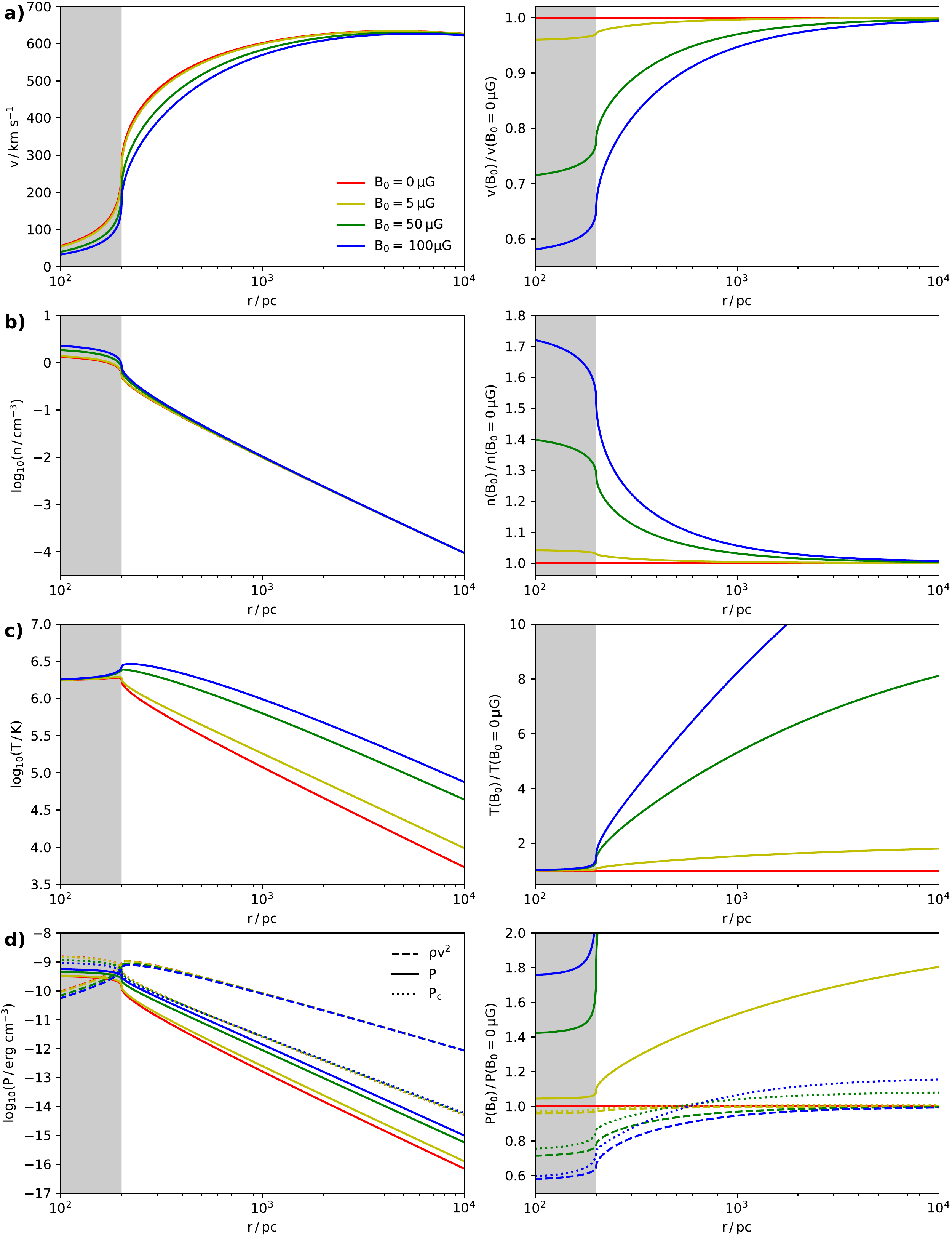}
\caption{\textbf{a)} Velocity, \textbf{b)} density, \textbf{c)} temperature and \textbf{d)} pressure profiles of CR-driven outflows are plotted on the left panels, where \(B_0\) is varied between \(0\,\mu \rm G\) (in red), \(5\,\mu \rm G\) (in yellow), \(50\,\mu \rm G\) (in green) and \(100\,\mu \rm G\) (in blue). The right panels are the corresponding residual plots normalised to the case where \(B_0=0\,\mu \rm G\). This shows that, when \(B\) is higher, more energy is transferred from  the CRs to heat up the thermal gas. This competes with the adiabatic cooling of the thermal gas during the wind expansion, and causes the pressure gradient in the wind to be lower, leading to a lower flow velocity.}
\label{fig:B_0}
\end{figure*}

\begin{figure*}
\includegraphics[width=2\columnwidth]{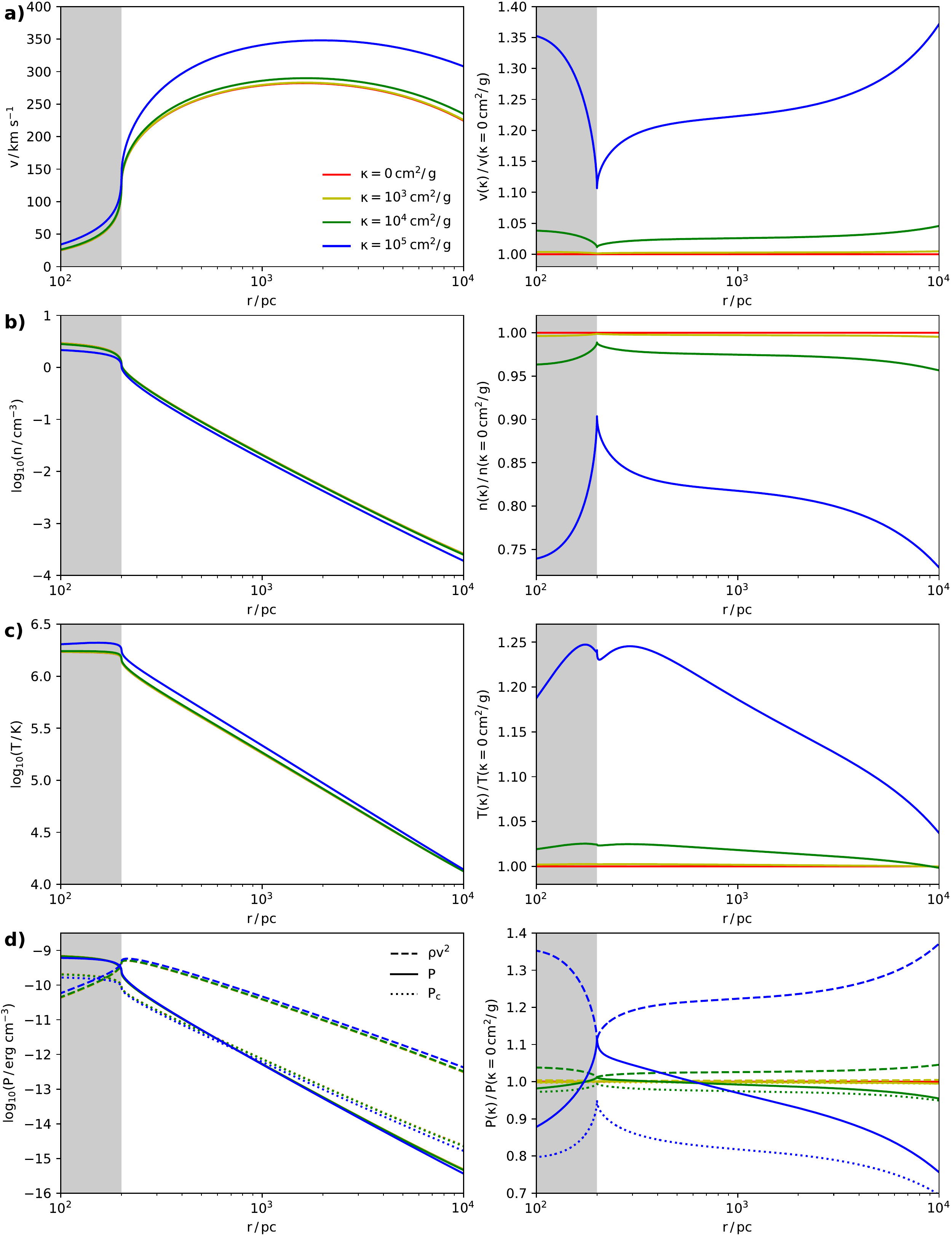}
\caption{\textbf{a)} Velocity, \textbf{b)} density, \textbf{c)} temperature and \textbf{d)} pressure profiles of radiation-driven outflows are plotted on the left panels, where \(\kappa\) is varied between \(0\,\rm cm^2/g\) (in red), \(10^3\,\rm cm^2/g\) (in yellow), \(10^4\,\rm cm^2/g\) (in green) and \(10^5\,\rm cm^2/g\) (in blue). The right panels are the corresponding residual plots normalised to the case where \(\kappa=0\,\rm cm^2/g\). When \(\kappa\) is low, the wind is mostly driven by thermal pressure and CRs even though most of the energy is injected as radiation. Radiation is important only when \(\kappa\) is sufficiently high for the wind to be optically thick.}
\label{fig:kappa}
\end{figure*}

\begin{figure*}
\includegraphics[width=1.9\columnwidth]{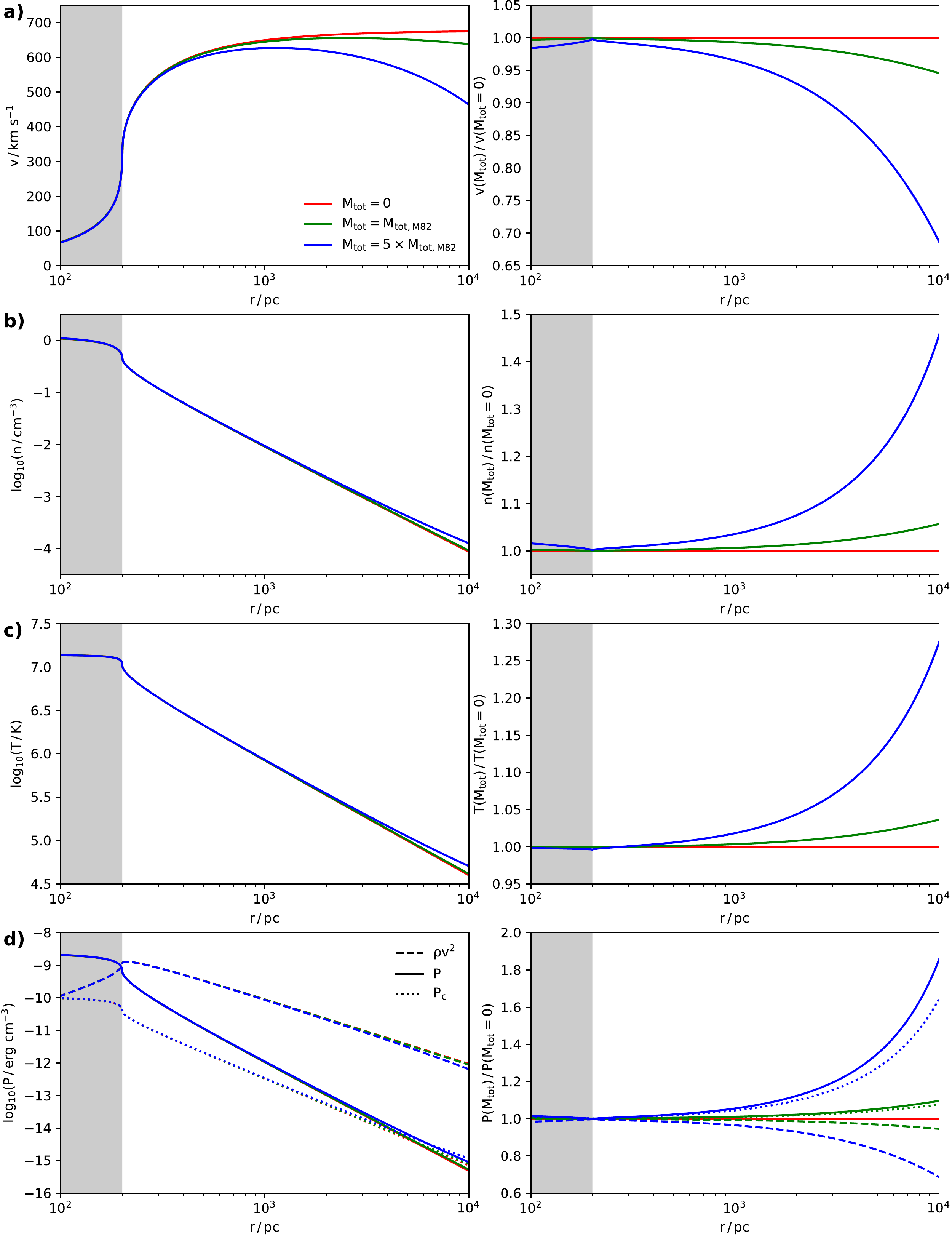}
\caption{\textbf{a)} Velocity, \textbf{b)} density, \textbf{c)} temperature and \textbf{d)} pressure profiles of galactic outflows are plotted on the left panels, where \(M_{\rm tot}\) is varied between \(0\) (in red), $M_{\rm tot,M82} = 5.54\times 10^{11}\text{M}_{\odot}$  (in green) and \(5\times M_{\rm tot,M82}\) (in blue). The right panels are the corresponding residual plots normalised to the case where \(M_{\rm tot}=0\). This shows that the effect of the gravitational potential is cumulative across \(r\). As such it is unimportant at low galactocentric distance but, if the galaxy is massive enough, the galactic wind can be significantly impeded or even bound by the gravitational potential.
}
\label{fig:G}
\end{figure*}

The magnetic field strength governs the degree to which the CR and thermal components of an outflow are coupled, as is explored in Fig.~\ref{fig:B_0} where we show the velocity, density, temperature and pressure profiles for outflows where different magnetic field strengths have been adopted in panels \textbf{a}, \textbf{b}, \textbf{c} and \textbf{d}, respectively. On the right, we show the corresponding residual plots, which are normalised to the case where \(B_0=0\) (the red line).
Note that the green curve (\(B_0=50\ \mu \rm G\)) is equivalent to the CR-driven outflow shown in Fig.~\ref{fig:mechanism}. 
We see that, 
among the HD variables, the flow temperature 
is the most sensitive one to the magnetic field strength. When \(B_0=0\ \mu \rm G\), the wind temperature at \(r=1\ \rm kpc\) would be around 8 times lower compared to the case where \(B_0=100\ \mu \rm G\). 
This is because the energy from the CRs is transferred to the thermal gas by two different channels, via thermal energy and via the bulk kinetic energy as quantified by equation~\ref{eq:CRtransfer}. $v_{\rm A}$ is proportional to the magnetic field strength, so when \(B_0=0\ \mu \rm G\), then $v_{\rm A} = 0$ and the CRs cannot couple with a magnetic field to facilitate the transfer of thermal energy to the wind. As such, they cannot heat the wind fluid via the damping of Alfv\'{e}n waves which they would otherwise excite, and the wind is colder as a result.

We find that stronger magnetic fields also lead to lower outflow velocities at all galactocentric distance, although this is especially pronounced slightly above the starburst radius where CR heating via the streaming instability is strongest (see panel \textbf{c}).  
Greater coupling between the CR component of the wind and the thermal component leads to more powerful energy transfer rates and higher temperatures. However this process particularly elevates the temperature just above the starburst radius, causing a reduction in the thermal pressure gradient. The result is a lower driving effect that yields a slower flow velocity (although the terminal wind velocity at \(r\to\infty\) would not be affected).
These results are consistent with the profiles for thermal pressure and ram pressure in panel \textbf{d}.
At lower altitudes, within the starburst radius, stronger magnetic fields yield a lower CR pressure (cf. equation \ref{eq:CRinitPc}).
This effect can be understood in terms of the Alfv\'{e}n velocity, which is higher in stronger magnetic fields, allowing for a faster propagation of CRs away from their injection point in this region (where the bulk flow velocity $v$ is comparatively negligible).

\subsection{Opacity}
\label{sec:kappa}

The mean opacity  ($\kappa$) of an outflow affects how strongly the gas is driven by radiation. This is because it governs the rate at which radiation can be absorbed by the wind material, and how quickly momentum may be transferred. We explore the sensitivity of the HD quantities to variations in flow opacity in Fig.~\ref{fig:kappa} for a radiation-driven wind. A reference case with $\kappa = 0$ is plotted in red, which would represent a wind fully driven by thermal and CR pressure. We note that the green curve with \(\kappa=10^4\,\rm cm^2/g\) is equivalent to the radiation-driven outflow in Fig.~\ref{fig:mechanism}. It can be seen that, as \(\kappa\) increases, more energy and momentum is transferred from the radiation to the outflowing gas, and radiative heating becomes more powerful.
This is demonstrated in panels \textbf{a} and \textbf{c}, where the wind velocity and temperature are indeed higher with increasing \(\kappa\) -- an effect that is also consistent with the ram pressure profile in panel \textbf{d}. Moreover, the wind density (panel~\textbf{b}) is lower with higher opacities, corresponding to the inverse relation between density and flow velocity. High values of \(\kappa\) similarly lead to lower CR pressure. This follows from the reduced CR density, caused by their advection in a faster wind. The impact on thermal pressure is more complicated: thermal energy is increased by higher opacities due to enhanced radiative heating. However, the thermal pressure usually
drops because the increase in wind velocity is more important.
An exception arises in the vicinity of \(r_{\rm sb}\), where the rise in wind temperature is maximised, but the increase in wind velocity is minimal.

We note that, although we have considered models where the energy injection is dominated by radiation, much of the available radiative power escapes from the (optically thin) wind -- unless unphysically high mean opacity values are adopted.
 In reality, some spectral lines may cause the wind to be optically thick to certain wavelengths. If these lines correspond to frequencies at which the irradiating spectrum contains substantial power, the transfer of energy may be sufficient to produce a line-driven outflow \citep[see, e.g.,][]{Lamers1999}.
This means that the properties of the line-driving mechanism cannot be explained using the mean opacity approach alone, with more careful consideration of the source spectrum and wind composition being required. Such matters are beyond the scope of the current paper, and are left to follow-up studies.

\begin{table*}
\centering
\begin{tabular}{*{7}{c}}
\midrule
Figure & Label & $T\,/\,$K & $n\,/\,\rm cm^{-3}$ & $P\,/\,\rm erg\ cm^{-3}$ & $P_{\rm C}\,/\,\rm erg\ cm^{-3}$ & $v\,/\,\rm km\ s^{-1}$ \\
\midrule
\ref{fig:mechanism} & therm & 1.382$\times10^7$ & 1.199 & 2.287$\times10^{-9}$ & 1.060$\times10^{-10}$ & 0 \\
\ref{fig:mechanism} & CR & 1.727$\times10^6$ & 2.061 & 4.914$\times10^{-10}$ & 1.293$\times10^{-9}$ & 0 \\
\ref{fig:mechanism} & rad & 1.727$\times10^6$ & 3.181 & 7.583$\times10^{-10}$ & 2.238$\times10^{-10}$ & 0 \\
\ref{fig:B_0} & $B_0=0\ \mu$G & 1.727$\times10^6$ & 1.452 & 3.461$\times10^{-10}$ & 1.730$\times10^{-9}$ & 0 \\
\ref{fig:B_0} & $B_0=5\ \mu$G & 1.727$\times10^6$ & 1.515 & 3.612$\times10^{-10}$ & 1.678$\times10^{-9}$ & 0 \\
\ref{fig:B_0} & $B_0=50\ \mu$G & 1.727$\times10^6$ & 2.061 & 4.914$\times10^{-10}$ & 1.293$\times10^{-9}$ & 0 \\
\ref{fig:B_0} & $B_0=100\ \mu$G & 1.727$\times10^6$ & 2.555 & 6.091$\times10^{-10}$ & 1.024$\times10^{-9}$ & 0 \\
\ref{fig:kappa} & $\kappa=0\ \rm cm^2/g$ & 1.727$\times10^6$ & 3.280 & 7.818$\times10^{-10}$ & 2.289$\times10^{-10}$ & 0 \\
\ref{fig:kappa} & $\kappa=10^3\ \rm cm^2/g$ & 1.727$\times10^6$ & 3.269 & 7.794$\times10^{-10}$ & 2.284$\times10^{-10}$ & 0 \\
\ref{fig:kappa} & $\kappa=10^4\ \rm cm^2/g$ & 1.727$\times10^6$ & 3.181 & 7.583$\times10^{-10}$ & 2.238$\times10^{-10}$ & 0 \\
\ref{fig:kappa} & $\kappa=10^5\ \rm cm^2/g$ & 1.727$\times10^6$ & 2.597 & 6.190$\times10^{-10}$ & 1.925$\times10^{-10}$ & 0 \\
\ref{fig:G} & $M_{\rm tot}=0$ & 1.382$\times10^7$ & 1.192 & 2.273$\times10^{-9}$ & 1.054$\times10^{-10}$ & 0 \\
\ref{fig:G} & $M_{\rm tot}=M_{\rm tot, M82}$ & 1.382$\times10^7$ & 1.199 & 2.287$\times10^{-9}$ & 1.060$\times10^{-10}$ & 0 \\
\ref{fig:G} & $M_{\rm tot}=5\times M_{\rm tot, M82}$ & 1.382$\times10^7$ & 1.230 & 2.346$\times10^{-9}$ & 1.081$\times10^{-10}$ & 0 \\
\midrule 
\end{tabular}
\caption{The initial conditions of the results in Section~\ref{sec:compare}. The values of the hydrodynamical variables at $r=0$ ($T$, $n$, $P$, $P_{\rm C}$, and $v$) are unique, which satisfy the boundary conditions $v=0$ at $r=0$ and $v=c_*$ at $r=r_{\rm sb}$. All values are rounded to 3 significant figures.}
\label{tab:init}
\end{table*}

\subsection{Gravitational potential}
\label{sec:G}

The gravitational potential of a massive galaxy influences the hydrodynamics of any ensuing outflow wind. If strong enough, it can even prevent the escape of wind material, keeping it gravitationally bound if wind speeds are sufficiently low. 
In this section, we assess the importance of the host galaxy mass \(M_{\rm tot}\) -- presumably dominated by the dark matter (DM) halo -- in impeding an outflow.
We assume a NFW-like DM halo with size and concentration specified by the reference values from Table~\ref{tab:param}, which are consistent with an M82-like galaxy~\citep{Oehm2017}. Our results are plotted in Fig.~\ref{fig:G}, with \(M_{\rm tot}=0\) (i.e. no gravitational effects) given by the red line, \(M_{\rm tot}=M_{\rm tot,M82}\) (green line) as a reference model, and \(M_{\rm tot}=5\times M_{\rm tot,M82}\) as an extreme comparison (blue line).
This shows, unsurprisingly, that larger halo masses will depress outflow wind velocities and, in extreme cases, can prevent the onset of a terminal velocity. The discrepancy with respect to the fiducial case becomes more substantial at larger flow altitudes where the cumulative work done by a wind in climbing out of the gravitational potential is greater. 
While the gravitational potential of M82 is not deep enough to change the velocity profile of its outflow significantly, the impact is much more apparent when the halo mass is increased fivefold, as shown in panel \textbf{a}. Lower flow velocities associated with deeper gravitational potentials will yield higher outflow densities (see panel \textbf{b}), as follows from mass continuity.
The pressure and temperature profiles (panel \textbf{d} and \textbf{c} respectively) behave in a similar way to the density profile. This is because the adiabatic expansion of the wind yields \(P\propto\rho^{\gamma_{\rm g}}\), \(P_{\rm C}\propto\rho^{\gamma_{\rm C}}\) and \(T\propto P/\rho\propto\rho^{\gamma_{\rm g}-1}\).
The halo concentration \(R_{\rm vir}/R_{\rm s}\) would affect an outflow in a similar manner to its total mass. This is because the size of the entire halo is determined by the virial radius \(R_{\rm vir}\), which is typically an order of magnitude greater than the visible scale of an outflow. Increasing the concentration will therefore amplify the density of the DM halo around the outflow and would steepen the gradient of the gravitational potential, making it harder for a wind to escape.

\subsection{Astrophysical implications}
\label{sec:implications}

Galactic outflows have an important role in the co-evolution of their host galaxies, circumgalactic environments and the intergalactic medium. They are expected to advect cosmic rays into circumgalactic and intergalactic space, where they may amplify intra-cluster/intergalactic magnetic fields through resistive generation as cosmic rays escape from galaxies during the cosmic dawn~\citep{Miniati2011, Beck2013, Lacki2015} or by driving the build-up of magnetohydrodynamical instabilities in weak seed magnetic fields~\citep{Bell2004, Miniati2011, Samui2018}. This may contribute to the growth of cosmological magnetic fields~\citep{Kronberg2016book, Durrer2013A&ARv} as might be detected by e.g. co-variant polarised radiative transfer methods~\citep{Chan2019}. Cosmic rays may also deposit energy to heat and ionise matter at high flow altitudes~\citep{Owen2019} possibly reaching into the circumgalactic and/or intergalactic medium, and this may contribute to the progression of cosmic pre-heating and reionisation~\citep[e.g.][]{Sazonov2015, Leite2017}. These advected cosmic rays may also provide pressure support around and between galaxies to balance against gravitational collapse in low-redshift clusters~\citep{Suto2013, Biffi2016} or affect the separation of bright neighbouring star-forming galaxies at high-redshift~\citep{Owen2019}.

Outflows are also able to transport matter and metals beyond the interior of a galaxy~\citep[e.g.][]{Songaila1997, Ellison2000, Bertone2005MNRASa, Aguirre2005}, and this has implications for subsequent star-forming episodes and cooling flows into and out of the interstellar/circumgalactic domain. Outflows modulate the mass-metallicity relation \citep[e.g.][]{Trem:04,Gall:05} and, along with the merger history, control the radial gradients of chemical composition in galaxies. 
The ability of outflows to realise their feedback potential is inextricably linked to their HD properties and their driving mechanisms. For instance, winds at high altitudes which would be the most important in transporting cosmic rays, hot gases and metals far into the intergalactic medium are thought to be driven predominantly by cosmic rays~\citep{Jacob2018}, while the faster winds that are able to transport energy and matter more quickly are instead more likely to be thermally-driven (cf. Fig.~\ref{fig:mechanism}). 
As such, being able to identify the nature of an outflow engine out to high-redshift in a way that simultaneously accounts for the possible contributions of multiple driving mechanisms is a very powerful probe of the extent to which feedback effects can be induced and sustained by observed galaxy populations. Providing access to outflow properties in the high-redshift Universe opens up new avenues to trace the evolution of galaxies, outflows and the action of (chemical and energetic) galaxy-scale feedback over cosmic time. A possible diagnostic to test this model would involve a battery of emission line measurements that are sensitive to differences in the physical properties of the outflowing gas, as it climbs out of the central regions of the galaxy. We will explore this methodology in a future paper.

\subsection{Additional remarks}
\label{sec:remark}

The HD models described in this work invoke a number of assumptions, chief among which is the flow being considered in a steady-state. This requires that the outflow timescale\footnote{This is defined as the timescale required for the wind flowing at its terminal velocity to traverse the distance from the base to the cap of the outflow.}~\citep[\(\sim\)20 Myr for M82, when the cap of the outflow is at 11.6 kpc, e.g.][]{Devine1999, Tsuru2007} is substantially shorter than the duration of the starburst episode driving it~\citep[\(\sim\)100s of Myr, e.g.][]{McQuinn2010, Hashimoto2018, McQuinn2018, Owen2019b}, a result supported by observations. This means that the mass and energy injection rates remain roughly steady, and a stationary outflow can therefore develop. A further requirement is that the radiative cooling timescale of the wind must be greater than the outflow timescale, otherwise the stationary flow solution is inhibited by run-away cooling and clumping before the outflow reaches its full extent. 
Such cooling/clumping would be expected in an M82-like outflow, and this is reflected in our results: we find that stationary solutions can generally be found only when radiative cooling is turned off for the cases considered in section \ref{sec:compare} and run-away cooling is artificially prevented. More generally, the stationary approximation for the outflow model enforces further assumptions regarding the micro-physics of the system. For example, the outflow is assumed to be inviscid, i.e. with no turbulence. This excludes micro-physics such as shock formation, wave propagation in an inhomogeneous wind medium and, perhaps most importantly, the multi-component multi-phase nature of a galactic outflow. These detailed matters can be taken into account in numerical HD simulations as demonstrated in, e.g.,  \cite{Scannapieco2010, Fujita2018}, but they come at the expense of the analytical simplicity and computational speed of the present approach.

We adopted solar metallicity abundances in our calculations, but in reality this may vary between starburst galaxies. Such an approximation would mainly affect the cooling rate (e.g. higher metallicity gases will typically undergo radiative cooling more quickly, \citealt{Sutherland1993ApJS}) and the mean molecular mass which determines temperature via the ideal gas law. We have used a mean molecular mass of $\mu = 1.4\,m_{\rm H}$ in line with \cite{Veilleux2005}, but other works have adopted either slightly different values of $\mu = 14/11\,m_{\rm H}$~\citep[e.g.][]{Wunsch2007}, or substantially different values of $\mu = 14/23\,m_{\rm H}$~\citep[e.g.][]{Silich2005}. A more rigorous treatment would take into account that the mean molecular mass is also dependent on the temperature, which governs the ionisation and recombination processes of electrons and ions in the wind (as modelled by the Saha equation).

\section{Conclusions}
\label{sec:conclusion}

In this work, we investigate the hydrodynamics of galactic outflows driven by three fundamental mechanisms. 
We adopt an analytical approach to determine the structures of thermal, radiative and cosmic ray-driven galactic outflows by solving the corresponding HD equations. 
We present a simple, phenomenological model which accounts for the contribution from all three mechanisms simultaneously. This model offers a generalised formulation for the study of the effects of all three driving mechanisms to the large scale hydrodynamics of galactic outflows.

For a starburst galaxy (such as the nearby M82), a thermally driven wind delivers the fastest and hottest outflow, the radiation driving mechanism is unable to develop a high velocity wind for realistic opacities, and cosmic rays yield less driving near the starburst nucleus compared to a thermally driven wind, becoming instead more important at larger galactocentric distances. 
A radiation-dominated outflow yields a slower wind that features a higher density throughout its extent compared to the other driving mechanisms. This makes the resulting outflow more susceptible to cooling and fragmentation, particularly at low galactocentric distance, where the densities are the highest.

We also assess the role of magnetic field, opacity and the gravitational potential due to the galactic mass in determining the subsequent properties of an outflow wind. 
We find that the magnetic field strength influences the coupling between a wind and the CR component, with stronger fields 
facilitating CR streaming to refocus much of their driving effect to higher altitudes.
Increasing the wind opacity leads to more effective radiative driving, but velocities competitive with the other two driving mechanisms can only be attained when the opacity is unphysically high.
Although the gravitational potential of M82 is not deep enough to impact the kinematics of its powerful outflow, the gravitational impact can be more significant in more massive galaxies which are strongly star-forming, such as the progenitors of the red nuggets found at cosmic noon.
  
  
 %

\section*{Acknowledgements} 
We thank Prof. Daisuke Kawata and Dr. Kuo-Chuan Pan for helpful discussions. 
BPBY and ERO thank the hospitality of the Institute of Astronomy, 
 National Tsing Hua University (NTHU), where part of this work was undertaken. 
BPBY's visit to NTHU was supported by the Ministry of Science and Technology of Republic of China (Taiwan) through grant 107-2112-M-007-032-MY3, and was hosted by Dr. Kuo-Chuan Pan. 
ERO's visit to NTHU was supported by the Ministry of Science and Technology of the Republic of China (Taiwan) through grants 105-2119-M-007-028-MY3 and 107-2628-M-007-003
 and was hosted by Prof. Albert Kong.  
ERO also acknowledges support by a UK Science and Technology Facilities Council PhD studentship. 
KW thanks the hospitality of Perimeter Institute where part of this work was carried out. 
Research at Perimeter Institute is supported in part by the Government of Canada through the Department of Innovation, Science and Economic Development Canada and by the Province of Ontario through the Ministry of Economic Development, Job Creation and Trade. 
This research has made use of NASA's Astrophysics Data Systems.



\bibliographystyle{mnras}
\bibliography{Brian2019MNRAS}

\begin{thebibliography}{}
\makeatletter
\relax
\def\mn@urlcharsother{\let\do\@makeother \do\$\do\&\do\#\do\^\do\_\do\%\do\~}
\def\mn@doi{\begingroup\mn@urlcharsother \@ifnextchar [ {\mn@doi@}
  {\mn@doi@[]}}
\def\mn@doi@[#1]#2{\def\@tempa{#1}\ifx\@tempa\@empty \href
  {http://dx.doi.org/#2} {doi:#2}\else \href {http://dx.doi.org/#2} {#1}\fi
  \endgroup}
\def\mn@eprint#1#2{\mn@eprint@#1:#2::\@nil}
\def\mn@eprint@arXiv#1{\href {http://arxiv.org/abs/#1} {{\tt arXiv:#1}}}
\def\mn@eprint@dblp#1{\href {http://dblp.uni-trier.de/rec/bibtex/#1.xml}
  {dblp:#1}}
\def\mn@eprint@#1:#2:#3:#4\@nil{\def\@tempa {#1}\def\@tempb {#2}\def\@tempc
  {#3}\ifx \@tempc \@empty \let \@tempc \@tempb \let \@tempb \@tempa \fi \ifx
  \@tempb \@empty \def\@tempb {arXiv}\fi \@ifundefined
  {mn@eprint@\@tempb}{\@tempb:\@tempc}{\expandafter \expandafter \csname
  mn@eprint@\@tempb\endcsname \expandafter{\@tempc}}}

\bibitem[\protect\citeauthoryear{{Aguirre}, {Schaye}, {Hernquist}, {Kay},
  {Springel}  \& {Theuns}}{{Aguirre} et~al.}{2005}]{Aguirre2005}
{Aguirre} A.,  {Schaye} J.,  {Hernquist} L.,  {Kay} S.,  {Springel} V.,
  {Theuns} T.,  2005, \mn@doi [\apjl] {10.1086/428498}, \href
  {http://adsabs.harvard.edu/abs/2005ApJ...620L..13A} {620, L13}

\bibitem[\protect\citeauthoryear{Ajiki et~al.,}{Ajiki
  et~al.}{2002}]{Ajiki2002ApJ}
Ajiki M.,  et~al., 2002, \apj, 576, L25

\bibitem[\protect\citeauthoryear{{Arribas, S.}, {Colina, L.}, {Bellocchi, E.},
  {Maiolino, R.}  \& {Villar-Martin, M.}}{{Arribas, S.}
  et~al.}{2014}]{Arribas2014A&A}
{Arribas, S.} {Colina, L.} {Bellocchi, E.} {Maiolino, R.}  {Villar-Martin, M.}
  2014, \mn@doi [\aap] {10.1051/0004-6361/201323324}, 568, A14

\bibitem[\protect\citeauthoryear{{Beck}, {Hanasz}, {Lesch}, {Remus}  \&
  {Stasyszyn}}{{Beck} et~al.}{2013}]{Beck2013}
{Beck} A.~M.,  {Hanasz} M.,  {Lesch} H.,  {Remus} R.-S.,   {Stasyszyn} F.~A.,
  2013, \mn@doi [\mnras] {10.1093/mnrasl/sls026}, \href
  {http://adsabs.harvard.edu/abs/2013MNRAS.429L..60B} {429, L60}

\bibitem[\protect\citeauthoryear{{Bell}}{{Bell}}{2004}]{Bell2004}
{Bell} A.~R.,  2004, \mn@doi [\mnras] {10.1111/j.1365-2966.2004.08097.x}, \href
  {http://adsabs.harvard.edu/abs/2004MNRAS.353..550B} {353, 550}

\bibitem[\protect\citeauthoryear{{Ben{\'{\i}}tez}, {Broadhurst}, {Frye},
  {Lidman}, {King}, {Meylan}  \& {Schneider}}{{Ben{\'{\i}}tez}
  et~al.}{2002}]{Benitez2002book}
{Ben{\'{\i}}tez} N.,  {Broadhurst} T.,  {Frye} B.,  {Lidman} C.,  {King} L.,
  {Meylan} G.,   {Schneider} P.,  2002, in {Gilfanov} M.,  {Sunyeav} R.,
  {Churazov} E.,  eds, Lighthouses of the Universe: The Most Luminous Celestial
  Objects and Their Use for Cosmology. p.~239

\bibitem[\protect\citeauthoryear{{Bertone}, {Stoehr}  \& {White}}{{Bertone}
  et~al.}{2005}]{Bertone2005MNRASa}
{Bertone} S.,  {Stoehr} F.,   {White} S.~D.~M.,  2005, \mn@doi [\mnras]
  {10.1111/j.1365-2966.2005.08772.x}, \href
  {http://adsabs.harvard.edu/abs/2005MNRAS.359.1201B} {359, 1201}

\bibitem[\protect\citeauthoryear{{Biffi} et~al.,}{{Biffi}
  et~al.}{2016}]{Biffi2016}
{Biffi} V.,  et~al., 2016, \mn@doi [\apj] {10.3847/0004-637X/827/2/112}, \href
  {http://adsabs.harvard.edu/abs/2016ApJ...827..112B} {827, 112}

\bibitem[\protect\citeauthoryear{Bland-Hawthorn, Veilleux  \&
  Cecil}{Bland-Hawthorn et~al.}{2007}]{Bland-Hawthorn2007APSS}
Bland-Hawthorn J.,  Veilleux S.,   Cecil G.,  2007, \mn@doi [\apss]
  {10.1007/s10509-007-9567-8}, 311, 87

\bibitem[\protect\citeauthoryear{Bordoloi et~al.,}{Bordoloi
  et~al.}{2011}]{Bordoloi2011ApJ}
Bordoloi R.,  et~al., 2011, \apj, 743, 10

\bibitem[\protect\citeauthoryear{Bordoloi, Rigby, Tumlinson, Bayliss, Sharon,
  Gladders  \& Wuyts}{Bordoloi et~al.}{2016}]{Bordoloi2016MNRAS}
Bordoloi R.,  Rigby J.~R.,  Tumlinson J.,  Bayliss M.~B.,  Sharon K.,  Gladders
  M.~G.,   Wuyts E.,  2016, \mn@doi [\mnras] {10.1093/mnras/stw449}, 458, 1891

\bibitem[\protect\citeauthoryear{{Cant{\'o}}, {Raga}  \&
  {Rodr{\'{\i}}guez}}{{Cant{\'o}} et~al.}{2000}]{Canto2000}
{Cant{\'o}} J.,  {Raga} A.~C.,   {Rodr{\'{\i}}guez} L.~F.,  2000, \mn@doi
  [\apj] {10.1086/308983}, \href
  {http://adsabs.harvard.edu/abs/2000ApJ...536..896C} {536, 896}

\bibitem[\protect\citeauthoryear{{Caprioli}}{{Caprioli}}{2012}]{Caprioli2012JCAP}
{Caprioli} D.,  2012, \mn@doi [\jcap] {10.1088/1475-7516/2012/07/038}, \href
  {http://adsabs.harvard.edu/abs/2012JCAP...07..038C} {7, 038}

\bibitem[\protect\citeauthoryear{{Cecil}, {Ferruit}  \& {Veilleux}}{{Cecil}
  et~al.}{2002a}]{Cecil2002RMxAACS}
{Cecil} G.,  {Ferruit} P.,   {Veilleux} S.,  2002a, Revista Mexicana de
  Astronomia y Astrofisica Conference Series, 13, 170

\bibitem[\protect\citeauthoryear{Cecil, Bland-Hawthorn  \& Veilleux}{Cecil
  et~al.}{2002b}]{Cecil2002ApJ}
Cecil G.,  Bland-Hawthorn J.,   Veilleux S.,  2002b, \apj, 576, 745

\bibitem[\protect\citeauthoryear{{Chan}, {Wu}, {On}, {Barnes}, {McEwen}  \&
  {Kitching}}{{Chan} et~al.}{2019}]{Chan2019}
{Chan} J.~Y.~H.,  {Wu} K.,  {On} A.~Y.~L.,  {Barnes} D.~J.,  {McEwen} J.~D.,
  {Kitching} T.~D.,  2019, \mn@doi [\mnras] {10.1093/mnras/sty3498}, \href
  {http://adsabs.harvard.edu/abs/2019MNRAS.484.1427C} {484, 1427}

\bibitem[\protect\citeauthoryear{{Chevalier} \& {Clegg}}{{Chevalier} \&
  {Clegg}}{1985}]{CC1985}
{Chevalier} R.~A.,  {Clegg} A.~W.,  1985, \mn@doi [\nat] {10.1038/317044a0},
  \href {http://adsabs.harvard.edu/abs/1985Natur.317...44C} {317, 44}

\bibitem[\protect\citeauthoryear{{Cooper}, {Bicknell}, {Sutherland}  \&
  {Bland-Hawthorn}}{{Cooper} et~al.}{2008}]{Cooper2008ApJ}
{Cooper} J.~L.,  {Bicknell} G.~V.,  {Sutherland} R.~S.,   {Bland-Hawthorn} J.,
  2008, \mn@doi [\apj] {10.1086/524918}, \href
  {http://adsabs.harvard.edu/abs/2008ApJ...674..157C} {674, 157}

\bibitem[\protect\citeauthoryear{{Dav{\'e}}}{{Dav{\'e}}}{2009}]{Dave2009}
{Dav{\'e}} R.,  2009, in {Jogee} S.,  {Marinova} I.,  {Hao} L.,   {Blanc}
  G.~A.,  eds,  Astronomical Society of the Pacific Conference Series Vol. 419,
  Galaxy Evolution: Emerging Insights and Future Challenges. p.~347 (\mn@eprint
  {arXiv} {0901.3149})

\bibitem[\protect\citeauthoryear{{Dermer} \& {Powale}}{{Dermer} \&
  {Powale}}{2013}]{Dermer2013A&A}
{Dermer} C.~D.,  {Powale} G.,  2013, \mn@doi [\aap]
  {10.1051/0004-6361/201220394}, \href
  {http://adsabs.harvard.edu/abs/2013A%26A...553A..34D} {553, A34}

\bibitem[\protect\citeauthoryear{{Devine} \& {Bally}}{{Devine} \&
  {Bally}}{1999}]{Devine1999}
{Devine} D.,  {Bally} J.,  1999, \mn@doi [\apj] {10.1086/306582}, \href
  {http://ukads.nottingham.ac.uk/abs/1999ApJ...510..197D} {510, 197}

\bibitem[\protect\citeauthoryear{{Dickinson}, {Papovich}, {Ferguson}  \&
  {Budav{\'a}ri}}{{Dickinson} et~al.}{2003}]{Dickinson2003}
{Dickinson} M.,  {Papovich} C.,  {Ferguson} H.~C.,   {Budav{\'a}ri} T.,  2003,
  \mn@doi [\apj] {10.1086/368111}, \href
  {http://adsabs.harvard.edu/abs/2003ApJ...587...25D} {587, 25}

\bibitem[\protect\citeauthoryear{{Dijkstra} \& {Loeb}}{{Dijkstra} \&
  {Loeb}}{2008}]{Dijkstra2008MNRAS}
{Dijkstra} M.,  {Loeb} A.,  2008, \mn@doi [\mnras]
  {10.1111/j.1365-2966.2008.13920.x}, \href
  {http://adsabs.harvard.edu/abs/2008MNRAS.391..457D} {391, 457}

\bibitem[\protect\citeauthoryear{{Durrer} \& {Neronov}}{{Durrer} \&
  {Neronov}}{2013}]{Durrer2013A&ARv}
{Durrer} R.,  {Neronov} A.,  2013, \mn@doi [\aapr] {10.1007/s00159-013-0062-7},
  \href {https://ui.adsabs.harvard.edu/abs/2013A&ARv..21...62D} {21, 62}

\bibitem[\protect\citeauthoryear{{Ellison}, {Songaila}, {Schaye}  \&
  {Pettini}}{{Ellison} et~al.}{2000}]{Ellison2000}
{Ellison} S.~L.,  {Songaila} A.,  {Schaye} J.,   {Pettini} M.,  2000, \mn@doi
  [\aj] {10.1086/301511}, \href
  {http://adsabs.harvard.edu/abs/2000AJ....120.1175E} {120, 1175}

\bibitem[\protect\citeauthoryear{{Ferland} et~al.,}{{Ferland}
  et~al.}{2017}]{Ferland2017}
{Ferland} G.~J.,  et~al., 2017, \rmxaa, \href
  {http://adsabs.harvard.edu/abs/2017RMxAA..53..385F} {53, 385}

\bibitem[\protect\citeauthoryear{{Fields}, {Olive}, {Cass{\'e}}  \&
  {Vangioni-Flam}}{{Fields} et~al.}{2001}]{Fields2001A&A}
{Fields} B.~D.,  {Olive} K.~A.,  {Cass{\'e}} M.,   {Vangioni-Flam} E.,  2001,
  \mn@doi [\aap] {10.1051/0004-6361:20010251}, \href
  {http://adsabs.harvard.edu/abs/2001A%26A...370..623F} {370, 623}

\bibitem[\protect\citeauthoryear{Frye, Broadhurst  \& Benitez}{Frye
  et~al.}{2002}]{Frye2002ApJ}
Frye B.,  Broadhurst T.,   Benitez N.,  2002, \apj, 568, 558

\bibitem[\protect\citeauthoryear{{Fujita} \& {Mac Low}}{{Fujita} \& {Mac
  Low}}{2018}]{Fujita2018}
{Fujita} A.,  {Mac Low} M.-M.,  2018, \mn@doi [\mnras] {10.1093/mnras/sty715},
  \href {https://ui.adsabs.harvard.edu/abs/2018MNRAS.477..531F} {477, 531}

\bibitem[\protect\citeauthoryear{{Gallazzi}, {Charlot}, {Brinchmann}, {White}
  \& {Tremonti}}{{Gallazzi} et~al.}{2005}]{Gall:05}
{Gallazzi} A.,  {Charlot} S.,  {Brinchmann} J.,  {White} S.~D.~M.,   {Tremonti}
  C.~A.,  2005, \mn@doi [\mnras] {10.1111/j.1365-2966.2005.09321.x}, \href
  {https://ui.adsabs.harvard.edu/abs/2005MNRAS.362...41G} {362, 41}

\bibitem[\protect\citeauthoryear{{Hashimoto} et~al.,}{{Hashimoto}
  et~al.}{2018}]{Hashimoto2018}
{Hashimoto} T.,  et~al., 2018, \mn@doi [\nat] {10.1038/s41586-018-0117-z},
  \href {http://adsabs.harvard.edu/abs/2018Natur.557..392H} {557, 392}

\bibitem[\protect\citeauthoryear{{Heckman}}{{Heckman}}{2003}]{Heckman2003}
{Heckman} T.~M.,  2003, in {Avila-Reese} V.,  {Firmani} C.,  {Frenk} C.~S.,
  {Allen} C.,  eds,  Revista Mexicana de Astronomia y Astrofisica Conference
  Series Vol. 17, Revista Mexicana de Astronomia y Astrofisica Conference
  Series. pp 47--55

\bibitem[\protect\citeauthoryear{{Helder} et~al.,}{{Helder}
  et~al.}{2009}]{Helder2009}
{Helder} E.~A.,  et~al., 2009, \mn@doi [Science] {10.1126/science.1173383},
  \href {http://adsabs.harvard.edu/abs/2009Sci...325..719H} {325, 719}

\bibitem[\protect\citeauthoryear{{Hoopes}, {Heckman}, {Strickland}  \&
  {Howk}}{{Hoopes} et~al.}{2003}]{Hoopes2003}
{Hoopes} C.~G.,  {Heckman} T.~M.,  {Strickland} D.~K.,   {Howk} J.~C.,  2003,
  \mn@doi [\apjl] {10.1086/379533}, \href
  {http://ukads.nottingham.ac.uk/abs/2003ApJ...596L.175H} {596, L175}

\bibitem[\protect\citeauthoryear{{Ipavich}}{{Ipavich}}{1975}]{Ipavich1975}
{Ipavich} F.~M.,  1975, \mn@doi [\apj] {10.1086/153397}, \href
  {http://adsabs.harvard.edu/abs/1975ApJ...196..107I} {196, 107}

\bibitem[\protect\citeauthoryear{{Jacob}, {Pakmor}, {Simpson}, {Springel}  \&
  {Pfrommer}}{{Jacob} et~al.}{2018}]{Jacob2018}
{Jacob} S.,  {Pakmor} R.,  {Simpson} C.~M.,  {Springel} V.,   {Pfrommer} C.,
  2018, \mn@doi [\mnras] {10.1093/mnras/stx3221}, \href
  {http://adsabs.harvard.edu/abs/2018MNRAS.475..570J} {475, 570}

\bibitem[\protect\citeauthoryear{{Klein}, {Wielebinski}  \& {Morsi}}{{Klein}
  et~al.}{1988}]{Klein1988}
{Klein} U.,  {Wielebinski} R.,   {Morsi} H.~W.,  1988, \aap, \href
  {http://adsabs.harvard.edu/abs/1988A%26A...190...41K} {190, 41}

\bibitem[\protect\citeauthoryear{{Kronberg}}{{Kronberg}}{2016}]{Kronberg2016book}
{Kronberg} P.~P.,  2016, {Cosmic Magnetic Fields}.
Cambridge University Press

\bibitem[\protect\citeauthoryear{{Lacki}}{{Lacki}}{2015}]{Lacki2015}
{Lacki} B.~C.,  2015, \mn@doi [\mnras] {10.1093/mnrasl/slu186}, \href
  {http://adsabs.harvard.edu/abs/2015MNRAS.448L..20L} {448, L20}

\bibitem[\protect\citeauthoryear{{Lamers} \& {Cassinelli}}{{Lamers} \&
  {Cassinelli}}{1999}]{Lamers1999}
{Lamers} H.~J.~G.~L.~M.,  {Cassinelli} J.~P.,  1999, {Introduction to Stellar
  Winds}.
Cambridge University Press

\bibitem[\protect\citeauthoryear{{Lehnert}, {Heckman}  \& {Weaver}}{{Lehnert}
  et~al.}{1999}]{Lehnert1999ApJ}
{Lehnert} M.~D.,  {Heckman} T.~M.,   {Weaver} K.~A.,  1999, \mn@doi [\apj]
  {10.1086/307762}, \href {http://adsabs.harvard.edu/abs/1999ApJ...523..575L}
  {523, 575}

\bibitem[\protect\citeauthoryear{{Leite}, {Evoli}, {D'Angelo}, {Ciardi}, {Sigl}
   \& {Ferrara}}{{Leite} et~al.}{2017}]{Leite2017}
{Leite} N.,  {Evoli} C.,  {D'Angelo} M.,  {Ciardi} B.,  {Sigl} G.,   {Ferrara}
  A.,  2017, \mn@doi [\mnras] {10.1093/mnras/stx805}, \href
  {http://adsabs.harvard.edu/abs/2017MNRAS.469..416L} {469, 416}

\bibitem[\protect\citeauthoryear{{Lemoine-Goumard}, {Renaud}, {Vink}, {Allen},
  {Bamba}, {Giordano}  \& {Uchiyama}}{{Lemoine-Goumard}
  et~al.}{2012}]{Lemoine2012A&A}
{Lemoine-Goumard} M.,  {Renaud} M.,  {Vink} J.,  {Allen} G.~E.,  {Bamba} A.,
  {Giordano} F.,   {Uchiyama} Y.,  2012, \mn@doi [\aap]
  {10.1051/0004-6361/201219896}, \href
  {http://adsabs.harvard.edu/abs/2012A%26A...545A..28L} {545, A28}

\bibitem[\protect\citeauthoryear{{Lynds} \& {Sandage}}{{Lynds} \&
  {Sandage}}{1963}]{Lynds1963}
{Lynds} C.~R.,  {Sandage} A.~R.,  1963, \mn@doi [\apj] {10.1086/147579}, \href
  {http://adsabs.harvard.edu/abs/1963ApJ...137.1005L} {137, 1005}

\bibitem[\protect\citeauthoryear{{Madau}, {Ferguson}, {Dickinson},
  {Giavalisco}, {Steidel}  \& {Fruchter}}{{Madau} et~al.}{1996}]{Madau1996}
{Madau} P.,  {Ferguson} H.~C.,  {Dickinson} M.~E.,  {Giavalisco} M.,  {Steidel}
  C.~C.,   {Fruchter} A.,  1996, \mn@doi [\mnras] {10.1093/mnras/283.4.1388},
  \href {http://adsabs.harvard.edu/abs/1996MNRAS.283.1388M} {283, 1388}

\bibitem[\protect\citeauthoryear{{Mannucci}, {Cresci}, {Maiolino}, {Marconi}
  \& {Gnerucci}}{{Mannucci} et~al.}{2010}]{Mannucci2010}
{Mannucci} F.,  {Cresci} G.,  {Maiolino} R.,  {Marconi} A.,   {Gnerucci} A.,
  2010, \mn@doi [\mnras] {10.1111/j.1365-2966.2010.17291.x}, \href
  {http://adsabs.harvard.edu/abs/2010MNRAS.408.2115M} {408, 2115}

\bibitem[\protect\citeauthoryear{{Mart{\'{\i}}n-Fern{\'a}ndez},
  {Jim{\'e}nez-Vicente}, {Zurita}, {Mediavilla}  \&
  {Castillo-Morales}}{{Mart{\'{\i}}n-Fern{\'a}ndez}
  et~al.}{2016}]{Martin-Fernandez2016}
{Mart{\'{\i}}n-Fern{\'a}ndez} P.,  {Jim{\'e}nez-Vicente} J.,  {Zurita} A.,
  {Mediavilla} E.,   {Castillo-Morales} {\'A}.,  2016, \mn@doi [\mnras]
  {10.1093/mnras/stw1048}, \href
  {http://ukads.nottingham.ac.uk/abs/2016MNRAS.461....6M} {461, 6}

\bibitem[\protect\citeauthoryear{Martin, Shapley, Coil, Kornei, Murray  \&
  Pancoast}{Martin et~al.}{2013}]{Martin2013ApJ}
Martin C.~L.,  Shapley A.~E.,  Coil A.~L.,  Kornei K.~A.,  Murray N.,
  Pancoast A.,  2013, \apj, 770, 41

\bibitem[\protect\citeauthoryear{{McKeith}, {Greve}, {Downes}  \&
  {Prada}}{{McKeith} et~al.}{1995}]{McKeith1995A&A}
{McKeith} C.~D.,  {Greve} A.,  {Downes} D.,   {Prada} F.,  1995, \aap, \href
  {http://adsabs.harvard.edu/abs/1995A%26A...293..703M} {293, 703}

\bibitem[\protect\citeauthoryear{{McQuinn} et~al.,}{{McQuinn}
  et~al.}{2010}]{McQuinn2010}
{McQuinn} K.~B.~W.,  et~al., 2010, \mn@doi [\apj] {10.1088/0004-637X/724/1/49},
  \href {http://adsabs.harvard.edu/abs/2010ApJ...724...49M} {724, 49}

\bibitem[\protect\citeauthoryear{{McQuinn}, {Skillman}, {Heilman}, {Mitchell}
  \& {Kelley}}{{McQuinn} et~al.}{2018}]{McQuinn2018}
{McQuinn} K.~B.~W.,  {Skillman} E.~D.,  {Heilman} T.~N.,  {Mitchell} N.~P.,
  {Kelley} T.,  2018, \mn@doi [\mnras] {10.1093/mnras/sty839}, \href
  {http://adsabs.harvard.edu/abs/2018MNRAS.477.3164M} {477, 3164}

\bibitem[\protect\citeauthoryear{{Melioli}, {de Gouveia Dal Pino}  \&
  {Geraissate}}{{Melioli} et~al.}{2013}]{Melioli2013}
{Melioli} C.,  {de Gouveia Dal Pino} E.~M.,   {Geraissate} F.~G.,  2013,
  \mn@doi [\mnras] {10.1093/mnras/stt126}, \href
  {http://ukads.nottingham.ac.uk/abs/2013MNRAS.430.3235M} {430, 3235}

\bibitem[\protect\citeauthoryear{{Miniati} \& {Bell}}{{Miniati} \&
  {Bell}}{2011}]{Miniati2011}
{Miniati} F.,  {Bell} A.~R.,  2011, \mn@doi [\apj]
  {10.1088/0004-637X/729/1/73}, \href
  {http://adsabs.harvard.edu/abs/2011ApJ...729...73M} {729, 73}

\bibitem[\protect\citeauthoryear{{Morlino} \& {Caprioli}}{{Morlino} \&
  {Caprioli}}{2012}]{Morlino2012A&A}
{Morlino} G.,  {Caprioli} D.,  2012, \mn@doi [\aap]
  {10.1051/0004-6361/201117855}, \href
  {http://adsabs.harvard.edu/abs/2012A%26A...538A..81M} {538, A81}

\bibitem[\protect\citeauthoryear{{Nath} \& {Silk}}{{Nath} \&
  {Silk}}{2009}]{Nath2009MNRAS}
{Nath} B.~B.,  {Silk} J.,  2009, \mn@doi [\mnras]
  {10.1111/j.1745-3933.2009.00670.x}, \href
  {http://adsabs.harvard.edu/abs/2009MNRAS.396L..90N} {396, L90}

\bibitem[\protect\citeauthoryear{{Navarro}, {Frenk}  \& {White}}{{Navarro}
  et~al.}{1996}]{Navarro1996}
{Navarro} J.~F.,  {Frenk} C.~S.,   {White} S.~D.~M.,  1996, \mn@doi [\apj]
  {10.1086/177173}, \href {http://adsabs.harvard.edu/abs/1996ApJ...462..563N}
  {462, 563}

\bibitem[\protect\citeauthoryear{{Oehm}, {Thies}  \& {Kroupa}}{{Oehm}
  et~al.}{2017}]{Oehm2017}
{Oehm} W.,  {Thies} I.,   {Kroupa} P.,  2017, \mn@doi [\mnras]
  {10.1093/mnras/stw3381}, \href
  {https://ui.adsabs.harvard.edu/\#abs/2017MNRAS.467..273O} {467, 273}

\bibitem[\protect\citeauthoryear{{Ohyama} et~al.,}{{Ohyama}
  et~al.}{2002}]{Ohyama2002PASJ}
{Ohyama} Y.,  et~al., 2002, \mn@doi [\pasj] {10.1093/pasj/54.6.891}, \href
  {http://adsabs.harvard.edu/abs/2002PASJ...54..891O} {54, 891}

\bibitem[\protect\citeauthoryear{{Owen}, {Jin}, {Wu}  \& {Chan}}{{Owen}
  et~al.}{2019a}]{Owen2019}
{Owen} E.~R.,  {Jin} X.,  {Wu} K.,   {Chan} S.,  2019a, \mn@doi [\mnras]
  {10.1093/mnras/stz060}, \href
  {http://adsabs.harvard.edu/abs/2019MNRAS.484.1645O} {484, 1645}

\bibitem[\protect\citeauthoryear{{Owen}, {Wu}, {Jin}, {Surajbali}  \&
  {Kataoka}}{{Owen} et~al.}{2019b}]{Owen2019b}
{Owen} E.~R.,  {Wu} K.,  {Jin} X.,  {Surajbali} P.,   {Kataoka} N.,  2019b,
  \mn@doi [\aap] {10.1051/0004-6361/201834350}, \href
  {https://ui.adsabs.harvard.edu/abs/2019A&A...626A..85O} {626, A85}

\bibitem[\protect\citeauthoryear{Press, Teukolsky, Vetterling  \&
  Flannery}{Press et~al.}{2007}]{Press2007}
Press W.~H.,  Teukolsky S.~A.,  Vetterling W.~T.,   Flannery B.~P.,  2007,
  Numerical Recipes 3rd Edition: The Art of Scientific Computing, 3 edn.
Cambridge University Press: New York, NY

\bibitem[\protect\citeauthoryear{{Raymond}, {Cox}  \& {Smith}}{{Raymond}
  et~al.}{1976}]{Raymond1976}
{Raymond} J.~C.,  {Cox} D.~P.,   {Smith} B.~W.,  1976, \mn@doi [\apj]
  {10.1086/154170}, \href {http://adsabs.harvard.edu/abs/1976ApJ...204..290R}
  {204, 290}

\bibitem[\protect\citeauthoryear{{Rodr{\'{\i}}guez-Gonz{\'a}lez}, {Cant{\'o}},
  {Esquivel}, {Raga}  \& {Vel{\'a}zquez}}{{Rodr{\'{\i}}guez-Gonz{\'a}lez}
  et~al.}{2007}]{Rodriguez2007}
{Rodr{\'{\i}}guez-Gonz{\'a}lez} A.,  {Cant{\'o}} J.,  {Esquivel} A.,  {Raga}
  A.~C.,   {Vel{\'a}zquez} P.~F.,  2007, \mn@doi [\mnras]
  {10.1111/j.1365-2966.2007.12167.x}, \href
  {http://adsabs.harvard.edu/abs/2007MNRAS.380.1198R} {380, 1198}

\bibitem[\protect\citeauthoryear{Rubin, Prochaska, Koo, Phillips, Martin  \&
  Winstrom}{Rubin et~al.}{2014}]{Rubin2014ApJ}
Rubin K. H.~R.,  Prochaska J.~X.,  Koo D.~C.,  Phillips A.~C.,  Martin C.~L.,
  Winstrom L.~O.,  2014, \apj, 794, 156

\bibitem[\protect\citeauthoryear{Rupke, Veilleux  \& Sanders}{Rupke
  et~al.}{2005a}]{Rupke2005ApJS-a}
Rupke D.~S.,  Veilleux S.,   Sanders D.~B.,  2005a, \apjs, 160, 87

\bibitem[\protect\citeauthoryear{Rupke, Veilleux  \& Sanders}{Rupke
  et~al.}{2005b}]{Rupke2005ApJS-b}
Rupke D.~S.,  Veilleux S.,   Sanders D.~B.,  2005b, \apjs, 160, 115

\bibitem[\protect\citeauthoryear{{Samui}, {Subramanian}  \& {Srianand}}{{Samui}
  et~al.}{2010}]{Samui2010}
{Samui} S.,  {Subramanian} K.,   {Srianand} R.,  2010, \mn@doi [\mnras]
  {10.1111/j.1365-2966.2009.16099.x}, \href
  {http://adsabs.harvard.edu/abs/2010MNRAS.402.2778S} {402, 2778}

\bibitem[\protect\citeauthoryear{{Samui}, {Subramanian}  \& {Srianand}}{{Samui}
  et~al.}{2018}]{Samui2018}
{Samui} S.,  {Subramanian} K.,   {Srianand} R.,  2018, \mn@doi [\mnras]
  {10.1093/mnras/sty287}, \href
  {http://adsabs.harvard.edu/abs/2018MNRAS.476.1680S} {476, 1680}

\bibitem[\protect\citeauthoryear{{Sazonov} \& {Sunyaev}}{{Sazonov} \&
  {Sunyaev}}{2015}]{Sazonov2015}
{Sazonov} S.,  {Sunyaev} R.,  2015, \mn@doi [\mnras] {10.1093/mnras/stv2255},
  \href {http://adsabs.harvard.edu/abs/2015MNRAS.454.3464S} {454, 3464}

\bibitem[\protect\citeauthoryear{{Scannapieco} \& {Br{\"u}ggen}}{{Scannapieco}
  \& {Br{\"u}ggen}}{2010}]{Scannapieco2010}
{Scannapieco} E.,  {Br{\"u}ggen} M.,  2010, \mn@doi [\mnras]
  {10.1111/j.1365-2966.2010.16599.x}, \href
  {http://adsabs.harvard.edu/abs/2010MNRAS.405.1634S} {405, 1634}

\bibitem[\protect\citeauthoryear{{Sharma} \& {Nath}}{{Sharma} \&
  {Nath}}{2013}]{Sharma2013}
{Sharma} M.,  {Nath} B.~B.,  2013, \mn@doi [\apj] {10.1088/0004-637X/763/1/17},
  \href {http://adsabs.harvard.edu/abs/2013ApJ...763...17S} {763, 17}

\bibitem[\protect\citeauthoryear{{Shopbell} \& {Bland-Hawthorn}}{{Shopbell} \&
  {Bland-Hawthorn}}{1998}]{Shopbell1998}
{Shopbell} P.~L.,  {Bland-Hawthorn} J.,  1998, \mn@doi [\apj] {10.1086/305108},
  \href {http://adsabs.harvard.edu/abs/1998ApJ...493..129S} {493, 129}

\bibitem[\protect\citeauthoryear{{Silich}, {Tenorio-Tagle}  \&
  {Rodr{\'{\i}}guez-Gonz{\'a}lez}}{{Silich} et~al.}{2004}]{Silich2004}
{Silich} S.,  {Tenorio-Tagle} G.,   {Rodr{\'{\i}}guez-Gonz{\'a}lez} A.,  2004,
  \mn@doi [\apj] {10.1086/421702}, \href
  {http://adsabs.harvard.edu/abs/2004ApJ...610..226S} {610, 226}

\bibitem[\protect\citeauthoryear{{Silich}, {Tenorio-Tagle}  \&
  {A{\~n}orve-Zeferino}}{{Silich} et~al.}{2005}]{Silich2005}
{Silich} S.,  {Tenorio-Tagle} G.,   {A{\~n}orve-Zeferino} G.~A.,  2005, \mn@doi
  [\apj] {10.1086/497532}, \href
  {http://adsabs.harvard.edu/abs/2005ApJ...635.1116S} {635, 1116}

\bibitem[\protect\citeauthoryear{{Silich}, {Bisnovatyi-Kogan}, {Tenorio-Tagle}
  \& {Mart{\'{\i}}nez-Gonz{\'a}lez}}{{Silich} et~al.}{2011}]{Silich2011}
{Silich} S.,  {Bisnovatyi-Kogan} G.,  {Tenorio-Tagle} G.,
  {Mart{\'{\i}}nez-Gonz{\'a}lez} S.,  2011, \mn@doi [\apj]
  {10.1088/0004-637X/743/2/120}, \href
  {http://adsabs.harvard.edu/abs/2011ApJ...743..120S} {743, 120}

\bibitem[\protect\citeauthoryear{{Songaila}}{{Songaila}}{1997}]{Songaila1997}
{Songaila} A.,  1997, \mn@doi [\apjl] {10.1086/311006}, \href
  {http://adsabs.harvard.edu/abs/1997ApJ...490L...1S} {490, L1}

\bibitem[\protect\citeauthoryear{{Strickland}, {Ponman}  \&
  {Stevens}}{{Strickland} et~al.}{1997}]{Strickland1997A&A}
{Strickland} D.~K.,  {Ponman} T.~J.,   {Stevens} I.~R.,  1997, \aap, \href
  {http://adsabs.harvard.edu/abs/1997A%26A...320..378S} {320, 378}

\bibitem[\protect\citeauthoryear{Strickland, Heckman, Weaver  \&
  Dahlem}{Strickland et~al.}{2000}]{Strickland2000AJ}
Strickland D.~K.,  Heckman T.~M.,  Weaver K.~A.,   Dahlem M.,  2000, \aj, 120,
  2965

\bibitem[\protect\citeauthoryear{{Strickland}, {Heckman}, {Weaver}, {Hoopes}
  \& {Dahlem}}{{Strickland} et~al.}{2002}]{Strickland2002}
{Strickland} D.~K.,  {Heckman} T.~M.,  {Weaver} K.~A.,  {Hoopes} C.~G.,
  {Dahlem} M.,  2002, \mn@doi [\apj] {10.1086/338889}, \href
  {http://ukads.nottingham.ac.uk/abs/2002ApJ...568..689S} {568, 689}

\bibitem[\protect\citeauthoryear{{Strong}, {Porter}, {Digel},
  {J{\'o}hannesson}, {Martin}, {Moskalenko}, {Murphy}  \& {Orlando}}{{Strong}
  et~al.}{2010}]{Strong2010ApJ}
{Strong} A.~W.,  {Porter} T.~A.,  {Digel} S.~W.,  {J{\'o}hannesson} G.,
  {Martin} P.,  {Moskalenko} I.~V.,  {Murphy} E.~J.,   {Orlando} E.,  2010,
  \mn@doi [\apjl] {10.1088/2041-8205/722/1/L58}, \href
  {http://adsabs.harvard.edu/abs/2010ApJ...722L..58S} {722, L58}

\bibitem[\protect\citeauthoryear{{Sugahara}, {Ouchi}, {Harikane}, {Bouch{\'e}},
  {Mitchell}  \& {Blaizot}}{{Sugahara} et~al.}{2019}]{Sugahara2019}
{Sugahara} Y.,  {Ouchi} M.,  {Harikane} Y.,  {Bouch{\'e}} N.,  {Mitchell}
  P.~D.,   {Blaizot} J.,  2019, arXiv e-prints, \href
  {http://adsabs.harvard.edu/abs/2019arXiv190403106S} {}

\bibitem[\protect\citeauthoryear{{Sutherland} \& {Dopita}}{{Sutherland} \&
  {Dopita}}{1993}]{Sutherland1993ApJS}
{Sutherland} R.~S.,  {Dopita} M.~A.,  1993, \mn@doi [\apjs] {10.1086/191823},
  \href {http://adsabs.harvard.edu/abs/1993ApJS...88..253S} {88, 253}

\bibitem[\protect\citeauthoryear{{Suto}, {Kawahara}, {Kitayama}, {Sasaki},
  {Suto}  \& {Cen}}{{Suto} et~al.}{2013}]{Suto2013}
{Suto} D.,  {Kawahara} H.,  {Kitayama} T.,  {Sasaki} S.,  {Suto} Y.,   {Cen}
  R.,  2013, \mn@doi [\apj] {10.1088/0004-637X/767/1/79}, \href
  {http://adsabs.harvard.edu/abs/2013ApJ...767...79S} {767, 79}

\bibitem[\protect\citeauthoryear{{Thompson}, {Fabian}, {Quataert}  \&
  {Murray}}{{Thompson} et~al.}{2015}]{Thompson2015}
{Thompson} T.~A.,  {Fabian} A.~C.,  {Quataert} E.,   {Murray} N.,  2015,
  \mn@doi [\mnras] {10.1093/mnras/stv246}, \href
  {http://adsabs.harvard.edu/abs/2015MNRAS.449..147T} {449, 147}

\bibitem[\protect\citeauthoryear{{Tremonti} et~al.,}{{Tremonti}
  et~al.}{2004}]{Trem:04}
{Tremonti} C.~A.,  et~al., 2004, \mn@doi [\apj] {10.1086/423264}, \href
  {https://ui.adsabs.harvard.edu/abs/2004ApJ...613..898T} {613, 898}

\bibitem[\protect\citeauthoryear{{Tsuru} et~al.,}{{Tsuru}
  et~al.}{2007}]{Tsuru2007}
{Tsuru} T.~G.,  et~al., 2007, \mn@doi [\pasj] {10.1093/pasj/59.sp1.S269}, \href
  {http://ukads.nottingham.ac.uk/abs/2007PASJ...59S.269T} {59, 269}

\bibitem[\protect\citeauthoryear{Uhlig, Pfrommer, Sharma, Nath, En{\ss}lin  \&
  Springel}{Uhlig et~al.}{2012}]{Uhlig2012MNRAS}
Uhlig M.,  Pfrommer C.,  Sharma M.,  Nath B.~B.,  En{\ss}lin T.~A.,   Springel
  V.,  2012, \mn@doi [\mnras] {10.1111/j.1365-2966.2012.21045.x}, 423, 2374

\bibitem[\protect\citeauthoryear{{Veilleux}, {Cecil}  \&
  {Bland-Hawthorn}}{{Veilleux} et~al.}{2005}]{Veilleux2005}
{Veilleux} S.,  {Cecil} G.,   {Bland-Hawthorn} J.,  2005, \mn@doi [\araa]
  {10.1146/annurev.astro.43.072103.150610}, \href
  {http://adsabs.harvard.edu/abs/2005ARA%26A..43..769V} {43, 769}

\bibitem[\protect\citeauthoryear{{Wang} \& {Fields}}{{Wang} \&
  {Fields}}{2018}]{Wang2018MNRAS}
{Wang} X.,  {Fields} B.~D.,  2018, \mn@doi [\mnras] {10.1093/mnras/stx2917},
  \href {http://adsabs.harvard.edu/abs/2018MNRAS.474.4073W} {474, 4073}

\bibitem[\protect\citeauthoryear{{W{\"u}nsch}, {Silich}, {Palou{\v{s}}}  \&
  {Tenorio-Tagle}}{{W{\"u}nsch} et~al.}{2007}]{Wunsch2007}
{W{\"u}nsch} R.,  {Silich} S.,  {Palou{\v{s}}} J.,   {Tenorio-Tagle} G.,  2007,
  \mn@doi [\aap] {10.1051/0004-6361:20077282}, \href
  {https://ui.adsabs.harvard.edu/abs/2007A&A...471..579W} {471, 579}

\makeatother
\end{thebibliography}





\bsp	
\label{lastpage}
\end{document}